\documentclass[11pt,a4paper]{article}
\pdfoutput=1  
\usepackage{jcappub} 
\usepackage{newfloat} 
\usepackage{braket}
\usepackage{subcaption} 
\usepackage{lineno}
\usepackage[normalem]{ulem}

\makeatletter
\renewcommand*\env@matrix[1][\arraystretch]{%
  \edef\arraystretch{#1}%
  \hskip -\arraycolsep
  \let\@ifnextchar\new@ifnextchar
  \array{*\c@MaxMatrixCols c}}
\makeatother

\providecommand{\eqn}[1]{eq.~(\ref{eqn:#1})}

\providecommand{\tab}[1]{table~\ref{tab:#1}}
\providecommand{\fig}[1]{figure~\ref{fig:#1}}
\providecommand{\app}[1]{appendix~\ref{app:#1}}
\providecommand{\sect}[1]{sec.~\ref{sec:#1}}
\providecommand{\ref}[1]{ref.~(\cite{#1})}

\providecommand{\vecsymbol}[1]{\ensuremath{\boldsymbol{#1}}}

\providecommand{\gv}{\vecsymbol{g}}

\providecommand{\xv}{\vecsymbol{x}}

\providecommand{\Neffgrpf}{\ensuremath{N_{\text{eff}}}}
\providecommand{\Neffisof}{\ensuremath{N^{\text{isof}}_{\text{eff}}}}
\providecommand{\Neffpur}{\ensuremath{N^{\text{pur}}_{\text{eff}}}}
\providecommand{\eps}{\varepsilon}
\providecommand{\epsv}{\vecsymbol{\eps}}

\providecommand{\chiv}{\vecsymbol{\chi}}

\title{Effects of overlapping sources on cosmic shear estimation: Statistical sensitivity and pixel-noise bias}

\author[a, 1]{Javier Sanchez,\note{Corresponding author.}}
\author[b]{Ismael Mendoza,}
\author[c]{David P. Kirkby,}
\author[d]{and Patricia R. Burchat}
\author{for the LSST Dark Energy Science Collaboration}
\affiliation[a]{Fermi National Accelerator Laboratory, Batavia IL 60510, USA}
\affiliation[b]{Department of Physics, 
University of Michigan, 
Ann Arbor, MI 48109, USA}
\affiliation[c]{Department of Physics and Astronomy, University of California, Irvine, CA 92697, USA}
\affiliation[d]{Kavli Institute for Particle Astrophysics and Cosmology, Department of Physics, Stanford University, Stanford, CA 94305, USA}
\emailAdd{jsanch87@fnal.gov}
\emailAdd{imendoza@umich.edu}
\emailAdd{dkirkby@uci.edu}
\emailAdd{burchat@stanford.edu}

\abstract{The next generation of dark-energy imaging surveys -- so called ``Stage-IV'' surveys, such as that of the Rubin Observatory Legacy Survey of Space and Time (LSST) -- will cross a threshold in the number density of detected sources on the sky that requires qualitatively different image analysis and measurement techniques compared to the current generation of Stage-III surveys. 
In Stage-IV surveys, a significant amount of the cosmologically useful information is due to sources whose images overlap with those of other sources on the sky.   
We focus on the weak gravitational lensing probe, for which we expect the largest impact since the cosmic shear signal is primarily encoded in the estimated shapes of observed galaxies and thus directly impacted by overlaps. 
We introduce a framework based on the Fisher formalism to analyze the effect of the overlapping sources (``blending'') on the estimation of cosmic shear. This method gives concrete predictions for the minimum loss of information due to noise and blending for any choice of ``deblending" scheme and shape-measurement algorithm.
Our studies account for undetected sources but do not address their full effects and biases they may introduce.

We use simulated images and predict this impact of blending for three surveys: 
the Dark Energy Survey (DES), 
the Hyper-Suprime Cam Subaru Strategic Program (HSC-SSP), 
and the Rubin LSST.
Our methodology successfully estimates the statistical sensitivity to weak lensing for DES and HSC early results. 
For LSST, we present the expected loss in statistical sensitivity  for the ten-year survey due to blending.  
We  find that for approximately $62\%$ of galaxies that are likely to be detected in full-depth LSST images, at least 1\% of the flux in their pixels is from overlapping sources. We also find that the statistical correlations between measures of overlapping galaxies and, to a much lesser extent (0.2\%) the higher shot noise level due to their presence, decrease the effective number density of galaxies, $\Neffgrpf$, by $\sim 18\%$. We calculate an upper limit on $\Neffgrpf$ of $39.4$ galaxies per arcmin$^2$ in $r$ band. 
We study the impact of stars on $\Neffgrpf$ as a function of stellar density and illustrate the diminishing returns of extending the survey into lower Galactic latitudes.

We extend the simulation-based Fisher formalism to  predict the expected increase in pixel-noise bias due to blending for maximum-likelihood (ML) shape estimators.   
We find that noise bias depends sensitively on the particular shape estimator and measure of ensemble-average shape that is used, and properties of the galaxy that include redshift-dependent quantities such as size and luminosity.

The source code for these studies is available online.\footnote{The documented software developed for the catalog-level studies are available in the open-source LSST DESC github repository \url{https://github.com/LSSTDESC/WeakLensingDeblending}. The software for analyzing one or two galaxies with user-defined parameters is in the open-source github repository \url{https://github.com/ismael-mendoza/ShapeMeasurementFisherFormalism}.}
}
\arxivnumber{2103.02078}
\begin{document}
\vspace*{-\headsep}\vspace*{\headheight}
\hfill FERMILAB-PUB-21-048-E\\
\maketitle
\flushbottom

\section{Introduction}

The fact that the night sky appears mostly dark to the human eye provides key insights into the finite lifetime of luminous sources and the dilution effects of an expanding universe. 
On the other hand, the night sky is almost uniformly bright at the microwave frequencies associated with the cosmic microwave background~\cite{2014A&A...571A...1P}.
Fundamentally different techniques have been developed for surveying isolated sources at optical frequencies and a uniformly illuminated sky at microwave frequencies. 
The next generation of Stage-IV dark-energy imaging surveys~\cite{2013arXiv1309.5380W} will cross a threshold in the number density of detected sources on the sky that requires qualitatively different techniques than the current generation of Stage-III surveys. 
In Stage-IV surveys, a significant amount of the cosmologically useful information is due to sources whose images overlap with those of other sources on the sky. 
In this paper, we investigate two impacts of overlapping sources (statistical sensitivity and pixel-noise bias) for one particular dark-energy probe -- cosmic shear -- for both current and future dark-energy surveys.  

Current and planned dark-energy surveys exploit several complementary probes~\cite{2013PhR...530...87W}, including type-Ia supernovae, baryon acoustic oscillations, weak and strong gravitational lensing, the growth of large scale structure, and the abundance of galaxy clusters. 
All of these are adversely affected to some extent by overlapping images of galaxies and/or stars. 
For example, all imaging surveys rely on photometric methods to estimate redshift, but overlapping sources lead to more potential degeneracies and errors. 
In this paper, we focus on the weak gravitational lensing probe, where we expect the largest impact since the cosmic shear signal is primarily encoded in the estimated shapes of observed galaxies and thus directly impacted by overlaps. This topic has already been considered in refs.~\cite{2013MNRAS.434.2121C}, \cite{2015MNRAS.447.1746C}, and~\cite{2016ApJ...816...11D}. In this paper we introduce a framework based on image simulations and the Fisher formalism to analyze the effect of the overlapping sources (``blending'') on the estimation of cosmic shear, that is largely independent of the particular ``deblending" scheme used. 
We develop metrics for assessing two impacts of blending on galaxy shape measurement and cosmic shear estimation: loss of statistical sensitivity and increase in pixel-noise bias. 
For the study of statistical sensitivity, 
we consider the impact of two or more overlapping galaxies, and of overlapping stars and galaxies as a function of stellar density.
We use simulated images and address the impact of blending for three surveys: 
the Dark Energy Survey (DES)~\cite{2010JPhCS.259a2080S}, 
the Hyper-Suprime Cam Subaru Strategic Program (HSC-SSP)~\cite{2018PASJ...70S...4A}, 
and the Rubin Observatory Legacy Survey of Space and Time (LSST)~\cite{2019ApJ...873..111I}. 
For the noise bias study, we consider the impact of two or more overlapping galaxies for the LSST survey only.
 
We do not address systematic effects that depend on detailed modeling of the performance of \textit{specific} object detection or flux-sharing algorithms, or parameter estimation techniques;
however, we identify some of these potential systematic effects in our conclusions (sec.\,\ref{sec:conclusions}). 

An important effect that is not addressed by the Fisher formalism is the impact of so-called \textit{unrecognized} or \textit{ambiguous} blends~\cite{2016ApJ...816...11D}.  The Fisher formalism accounts for the covariance introduced by sources that may well be below the detection threshold for any detection algorithm, based on the true values of the source parameters. In real data, the sources that fall below the detection threshold will increase the overall noise in the inferred parameters for the detected sources, and may also bias the parameter estimates -- an effect that is not captured by the Fisher formalism. In that sense, our results may underestimate the expected uncertainty and bias on shape parameters.

We also do not address the impact of overlapping sources on photometry or photometric redshift measurements, or the systematic impact of star/galaxy misclassification. 

\section{Glossary and notation} 
To minimize ambiguities in notation and terminology used in this paper and in the literature, we define symbols in \tab{symbols} and terminology below, and then use these consistently in the text, tables, and figures.  

\begin{itemize}

\item We use the word ``blended'' to refer generically to objects whose surface brightness profiles overlap to some degree with at least one other object  when imaged by our telescopes  -- for example, two objects could be defined as blended if photons from both objects share the same pixel.  
The precise definition of when two objects are blended can depend on the specific analysis or the information that is available.

\item We use the phrases ``blending off'' and ``blending on'' to describe whether a set of  simulated objects is generated so that (i) each object is analyzed as if it has no overlaps, or (ii) correlations between overlapping (blended) objects in a group are included.

\item In different measures of signal-to-noise ratio  ($\nu$ in \tab{snr_defs}), the subscript ``isof'' refers to single isolated sources (with blending off) and the subscript ``grpf'' to groups of one or more sources (with blending on). 
In both subscripts, the ``f'' denotes that all source parameters are assumed to be free.

\item We define a ``deblender'' (or the corresponding verb ``to deblend'') as any algorithm that attempts to allocate the flux in a single pixel to two or more objects. 
This includes algorithms that attempt to remove flux due to neighboring objects before making a single-object measurement, and simultaneous model fitting to multiple objects.

\item We define ``detectable'' objects to be those \textit{true} objects that, based on the estimated signal-to-noise ratio of flux, are judged to be detectable with a detection algorithm (e.g., a peak-finding algorithm).

\item We define ``detected'' objects to be those \textit{true} objects that match an object found with a detection algorithm, such as SourceExtractor \cite{1996A&AS..117..393B}.

\end{itemize}

\begin{table}[htbp]
\begin{center}
\begin{tabular}{|llc|}
\hline
Symbol    &  Name  or brief definition  & More details  \\
\hline
\vspace{0.1pt} & & \\
$Q$    &  Symmetric 2nd-moment tensor of surface-brightness profile  & \eqn{Q} \\
$\sigma_{-}$    &  A measure of size: $\det(Q)^{1/4}$   & \eqn{size} \\
$\sigma_{+}$    &  A measure of size: $\left(\frac{Q_{11} + Q_{22}}{2}\right)^{1/2}$  & \eqn{size} \\	
$\epsv = \eps_1 + i \eps_2$    &  Complex ellipticity spinor, intrinsic galaxy shape   &   
\eqn{ellipticity} \\
$\epsv' = \eps'_1 + i \eps'_2$    &  Complex ellipticity spinor, sheared galaxy shape   & \eqn{lensing} \\
$\gv = g_1 + i g_2$    &  Complex reduced shear    & \sect{shapeshear} \\
$s_{ip}$    &  Expected signal for source $i$ in pixel $p$ ($e^-$/pixel)   & \sect{imagesim} \\
$s_{\rm min}$    &  Limiting surface brightness used in simulation ($e^-$/pixel)   & \sect{imagesim} \\
$b$    &  Mean sky background (electrons/pixel)   & \sect{imagesim} \\

$\theta_\alpha$    &  True value of parameters describing galaxy profile  & \sect{Fisherformalism} \\
$\hat{\theta}_\alpha$ & Maximum likelihood estimator (MLE) for parameter $\theta_\alpha$ &  \sect{noisebias} \\
$\mu_p$  & Total expected flux in pixel $p$  (electrons/pixel)  & \eqn{pixelflux} \\
$\mu_{p,\alpha}$, $\mu_{p,\alpha\beta}$  & Partial derivative with respect to $\theta_\alpha$, $\theta_\beta$: $\frac{\partial{\mu_p}}{\partial{\theta_\alpha}}$, $\frac{\partial^2{\mu_p}}{\partial{\theta_\alpha}\partial{\theta_\beta}}$  & sec.\,\ref{sec:statsensitivity} \\
$C$ & Covariance matrix associated with the pixel noise  & \sect{statsensitivity}  \\
$\cal{F}$ &  Fisher matrix for parameter estimators & \sect{statsensitivity}  \\
$\cal{C}$  & Covariance matrix associated with parameter estimators  & sec.\,\ref{sec:noisebias}  \\
$\cal{B}$   & Noise bias tensor  &
\eqn{bias_tensor} \\
$b_\alpha$ or $b(\theta_\alpha)$    &  Expected pixel-noise bias for parameter $\theta_\alpha$ for MLE  & 
\eqn{bias} \\
$\nu$    &  Signal-to-noise ratio (SNR) for an object   & \eqn{nu_gen}, \tab{snr_defs} \\
$\rho$    &  Purity of an object   & \eqn{rho} \\
$\sigma_{\text{I}}$ & Standard deviation of  intrinsic galaxy shape for detectable sources & sec.\,\ref{sec:Neff} \\
$\Neffgrpf$ & Effective number density of perfectly measured galaxy shapes & \sect{Neff} \\
$i$ and $r$    &  AB magnitudes in $i$ and $r$ band  & \\
\hline
\end{tabular}
\end{center}
\caption{
Brief definitions of symbols used in this paper. 
For more details, see the specific sections, equations, or table listed in the final column.}
\label{tab:symbols}
\end{table}

\section{Galaxy shape and shear estimators}

\label{sec:shapeshear}
We describe galaxy shapes using the complex ellipticity spinor defined as~\cite{2001PhR...340..291B}
\begin{equation}
\epsv = \frac{Q_{11} - Q_{22} + 2 i Q_{12}}{Q_{11} + Q_{22} + 2 \, \left(\det Q\right)^{1/2}} \;,
\label{eqn:ellipticity}
\end{equation}
where $\det$ means matrix determinant and
$Q_{ij}$ are the components of the symmetric second-moment tensor of the galaxy's surface-brightness profile $I(x_1,x_2)$ measured using angles $x_1$ and $x_2$ on the sky relative to the surface brightness centroid: 
\begin{equation}
Q_{ij} = \frac{\iint d x_1 d x_2 I(x_1,x_2) x_i x_j}{\iint d x_1 d x_2 I(x_1,x_2)}\,.
\label{eqn:Q}
\end{equation}

We can also associate $\epsv$ with an ellipse if we assume that we can write $I(\xv) = \det M \, p(|M \xv|)$, where $\left \vert \cdot \right \vert$ stands for the Euclidean norm, $p(r)$ is some radial profile, and $M$ an affine transform. In this case, the isophotes of $I(\xv)$ are all elliptical and self-similar, and have semi-major ($a$) and semi-minor ($b$) axes that satisfy
\begin{equation}
\epsv = \frac{a-b}{a+b}\, e^{2i\beta} \;,
\end{equation}
where $\beta$ is the ellipse position angle -- i.e., the counter-clockwise rotation of the semi-major axis from the $+\hat{x}_1$ direction.

We consider two measures of a galaxy's size based on its second moments: 
\begin{equation}
\sigma_{+} \equiv \left(\frac{Q_{11} + Q_{22}}{2}\right)^{1/2}  
\; , \quad
\sigma_{-} \equiv \left(\det Q\right)^{1/4} \; .
\label{eqn:size}
\end{equation}
These measures of size are related by $\sigma_- \le \sigma_+$ with equality for round ($\epsv = 0$) galaxies.\footnote{The measure $\sigma_+$ is smaller than the size $r_{\text{sec}}$ adopted in refs.\,\cite{2013MNRAS.434.2121C,2015MNRAS.447.1746C}: specifically,  $\sigma_+ = r_{\text{sec}}/\sqrt{2}$. 
We choose this normalization so that $\sigma_- = \sigma_+ = \sigma$ for a round Gaussian profile with RMS size $\sigma$.
The measure $\sigma_+$ is related to $T$, also adopted in \cite{2015MNRAS.447.1746C}, by $\sigma_{+} = \sqrt{T/2}$.
}

Weak gravitational lensing of the galaxy's image by intervening matter transforms $(x_1,x_2)$ to observed angles $(x_1',x_2')$ (measured relative to the observed centroid) via a linear transformation that is conventionally parametrized in terms of its inverse
\begin{equation}
\begin{pmatrix}
x_1\\
x_2
\end{pmatrix} = (1-\kappa)\begin{pmatrix}
1-g_1 & -g_2 \\
-g_2 & 1+g_1
\end{pmatrix}
\begin{pmatrix}
x_1'\\
x_2'
\end{pmatrix} \; ,
\end{equation}
where $\kappa > 0$ magnifies the image and the reduced shear components $g_1$ and $g_2$ determine the change in shape between the source and the image along and $45^\circ$ to the $\hat{x}_1$ direction, respectively. 

When $|\gv| < 1$ with $\gv \equiv g_1 + i g_2$, a galaxy's intrinsic shape $\epsv$ is transformed to an observed shape $\epsv'$ via (see, for example, ref.\,\cite{2005WLSchneider})
\begin{equation}
\epsv' = \frac{\epsv + \gv}{1 + \gv^{*} \epsv} \; .
\label{eqn:lensing}
\end{equation}
If the intrinsic ellipticities $\epsv$ are uniformly distributed in angle, then the mean ellipticity is an estimator for the reduced shear: 
\begin{equation}
\langle \epsv'\rangle = \gv\,.
\label{eqn:estimator}
\end{equation}

\section{Galaxy catalog}
\label{sec:galaxy_catalog}
We use one square degree of a simulated galaxy catalog prepared for the LSST Catalog Simulator, \textsc{CatSim}~\cite{2014SPIE.9150E..14C}.
Galaxies are modeled as the sum of disk (Sersic $n = 1$) and bulge (Sersic $n = 4$) components, with realistic distributions of galaxy size and shape. 

For the purpose of this study, the appearance of each galaxy in the catalog is described by nine parameters: an apparent magnitude $m$, a centroid position $(x,y)$ in right ascension and declination, a half-light radius $r$ and minor-to-major ellipse axis ratio $0 < q \le 1$ for its disk $(r_d,q_d)$ and bulge $(r_b,q_b)$ components, an ellipse position angle $\beta$, and a bulge-to-total flux ratio $f_b$. All parameters except for the magnitude are assumed to be the same in all wavelength bands. The centroid and position angle are assumed to be the same for the disk and bulge components, so the component second-moment tensors can be directly added to calculate the galaxy's combined intrinsic ellipticity, \eqn{ellipticity}, and size, \eqn{size}. About 1\% of the catalog galaxies include an AGN component that represents $\lesssim 10\%$ of the total flux, which we model as a point-like component. 

\begin{figure}[tbp]
\begin{center}
\includegraphics[width=6.0in]{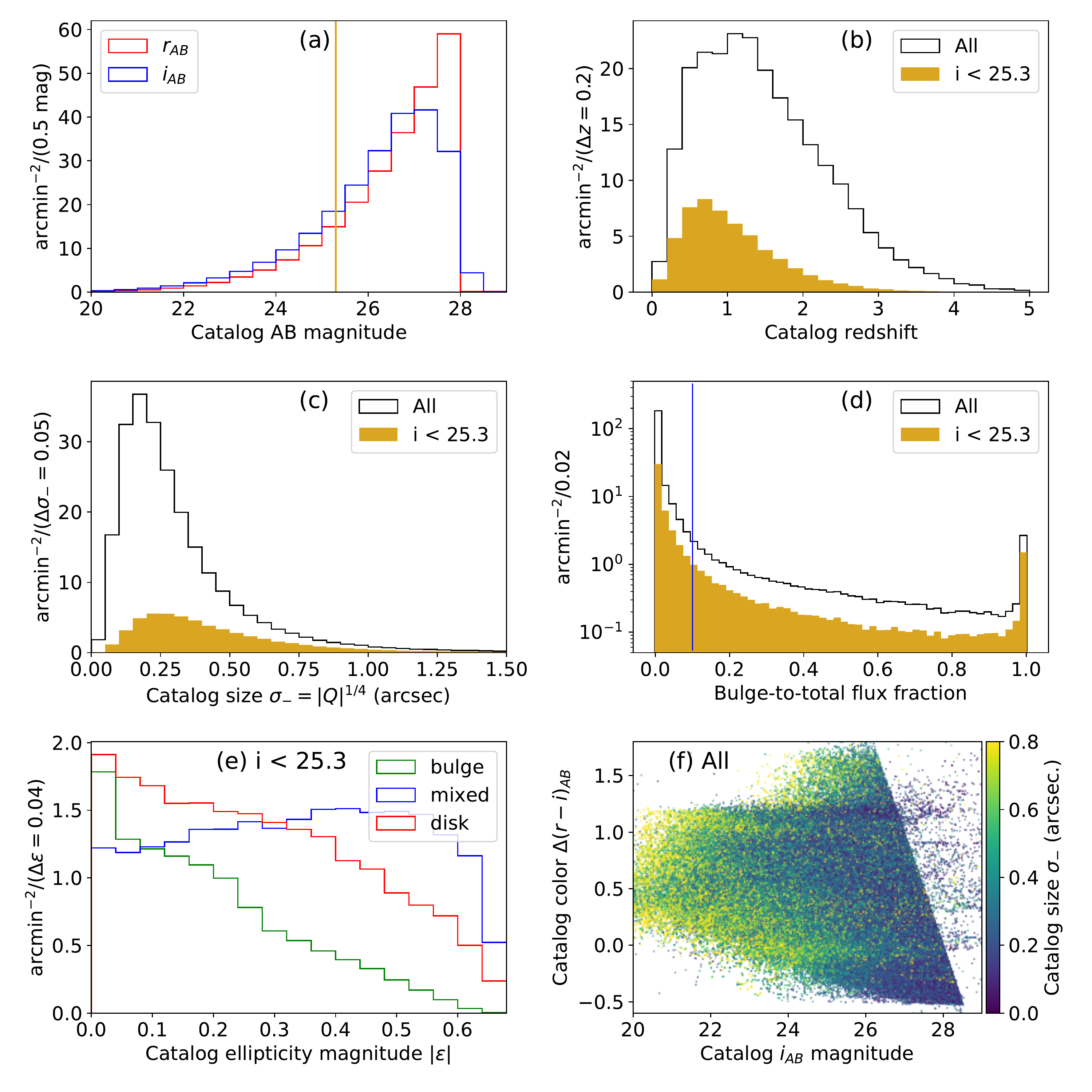}
\caption{Distributions of galaxy catalog quantities: (a) $i$- and $r$-band AB magnitudes, with a vertical line indicating the $i < 25.3$ cut that defines the LSST weak-lensing gold sample. Separately for all (outline histogram) and ``gold'' (filled histogram) catalog objects, the distributions of (b) redshifts, (c) galaxy sizes $\sigma_-$, and (d) fraction $f_b$ of total flux in a bulge component. The blue vertical line corresponds to $f_{b} = 0.1$. (e) Distribution of ellipticity magnitudes $|\epsv|$ for ``disk'' ($f_b = 0$), ``mixed'' ($0 < f_b < 0.1$), and ``bulge'' ($f_b \ge 0.1$) sources in the gold sample. (f) Correlations between $r-i$ color and $i$-band magnitude using a color scale to represent galaxy size $\sigma_-$.} 
\label{fig:catalog_plots}
\end{center}
\end{figure}

The one-square-degree catalog contains 858k galaxies (238 per square arcminute) out to the catalog limiting magnitude of $r < 28$. In order to evaluate the impact of cosmic variance and to ensure that the inferences using this one-square-degree catalog are not biased, we computed the standard deviation of the relative number density in patches of one square degree using 400 square degrees from the cosmoDC2 simulation~\citep{2019ApJS..245...26K}. We find that the standard deviation of the relative number density is $\sim 2\%$ so we expect our results to be accurate to this level. 
Restricting objects to the nominal LSST weak-lensing ``gold'' sample with $i < 25.3$~\cite{2009arXiv0912.0201L}, we find 53.7 galaxies per square arcminute, with a mean intrinsic ellipticity $\langle|\epsv|\rangle = 0.237$. 
Figure\,\ref{fig:catalog_plots} shows  distributions of galaxy catalog quantities.  We  observe that galaxy size and ellipticity are correlated with a galaxy's position in the color-magnitude plots; see~\fig{catalog_plots}(f).

\section{Star catalogs}
\label{sec:stellar_catalog}

We study the effects of overlapping galaxies and stars using five different one-square-degree fields in the \textsc{CatSim} catalog, chosen to cover a range of stellar densities.

We determine the distribution of stellar densities over the baseline LSST survey strategy~\cite{2019ApJ...873..111I} by selecting 200 random observations from a fiducial run of the LSST operations simulator \textsc{OpSim}\footnote{\url{http://www.lsst.org/scientists/simulations/opsim}}, restricted to galactic latitude greater than $15^{\circ}$ and a single-visit $5\sigma$ point-source depth of $r \ge 24.5$ to emulate the ``wide-fast-deep'' program. 
We then define low- and medium-density regions as those with stellar density $\rho_{*}$ at the 5th ($2.6$ stars/arcmin$^{2}$) and 50th ($6.3$ stars/arcmin$^{2}$) percentile for the 200 regions. 
We also include three additional observations with $\rho_{*}= 10.9$, 20.9, and 33.8 stars/arcmin$^{2}$ to study trends beyond the 50th percentile.

Stars are modeled as point sources in our image simulations.

\section{Image simulation}
\label{sec:imagesim}
We use \textsc{GalSim}~\cite{2015A&C....10..121R} to simulate the appearance of individual galaxies and stars in a ground-based imaging detector, accounting for the effects of a  point-spread function (PSF), pixelation, and overall detector response. We first define a detector pixel grid that maps catalog right ascension and declination into pixel coordinates for a fixed simulated telescope pointing. Next, we render each galaxy $i$, defined by its nine catalog parameters, into its own rectangular postage-stamp image with a signal $s_{ip}$ detected in pixel $p$. 
The rendering is performed semi-analytically using an efficient Fourier-transform convolution of its combined disk and bulge surface brightness (including a sub-pixel offset to the centroid position) with the assumed PSF and pixel response function.
Therefore, it is effectively a noise-free prediction of the mean expected signal. Pixel values $s_{ip}$ are calculated in units of detected electrons for the full simulated exposure, without any sky flux included.

The rectangular bounds of each galaxy's simulated postage stamp are chosen to enclose a limiting surface brightness isophote defined by
\begin{equation}
s_{ip} \ge s_{\rm min}\; ,
\label{eqn:threshold}
\end{equation}
where $s_{\rm min} \equiv \nu_{\text{pix}}\,b^{1/2}$,  $b$ is the mean sky level per pixel, and $\nu_{\text{pix}} = 0.05$ is the pixel signal-to-noise threshold used in the simulation. 
This approach leads to large postage stamps for bright sources but also allows us to study the effects of their overlaps with fainter sources. 
We drop from further consideration faint catalog objects whose maximum pixel value is below the limiting surface brightness defined in \eqn{threshold}.
We also zero all pixel values below the limiting surface brightness in our subsequent calculations, to ensure a consistent treatment of low-signal pixels that does not depend on how the limiting isophotes fit into their rectangular bounding boxes. The light profile for bright galaxies with bulge components is not described well at large distances from the galaxy centroid because the Sersic $n=4$ component has a relatively flat tail that eventually dominates the more rapidly damped $n=1$ tail. We truncate all extended sources at a radius of 30 arcsec to minimize the impact of these unphysical tails.

We simulate observations of the same catalog objects with three idealized ground-based surveys optimized for cosmic shear measurements: the Hyper-Suprime Cam Subaru Strategic Program (HSC-SSP), which consists of a 5-to-6 year program (300 nights) started in March 2014, with plans to map 1,400 sq.~deg.~in a wide survey, with a targeted depth of $i < 26.2$; the Dark Energy Survey (DES), which completed its 6-year (575 nights) science program between August 2013 and January 2019, and includes a 5,000 sq.~deg.~lensing survey to $i < 25.3$; and LSST, which will begin a 10-year survey in the early 2020s and include a 18,000 sq.~deg.~lensing survey to $r \simeq 27.2$ and $i \simeq 27.0$ ~\citep{2009arXiv0912.0201L}.
In the terminology of the Dark Energy Task Force~\cite{2006astro.ph..9591A}, these are Stage III (DES and HSC) and IV (LSST) dark energy projects.

The relevant simulation parameters describing each survey are its pixel size, nominal full-depth exposure time, level of sky background, quality of atmospheric seeing, and overall spectral throughput and detector response (``zero points'').\footnote{\noindent
In this study, we model only a mean sky level, mean seeing, etc., over an entire survey. 
This relative simplicity facilitates comparisons between surveys, and between band passes within or across surveys. 
In contrast, for a past LSST study \citep{2013MNRAS.434.2121C}, individual exposures were simulated with a range of sky background levels and atmospheric seeing, which requires a model for co-adding exposures.} 
We use the values given in \tab{simpar}. In this work, we focus on the $i$ and $r$ bands where the overall image quality is generally best for a lensing analysis, and ignore the small differences in the spectral responses of these bands between the three instruments being simulated.
\begin{table}[tbp]
\begin{center}
\setlength\tabcolsep{4.5pt}
\begin{tabular}{|lcccccccc|}
\hline
       & Effective   & Primary  & Pixel & Median   &   & Exp.  & Sky        & Atmos. \\
       & Area        & Diameter & Size  & Airmass   &   & Time  & Brightness & FWHM   \\
Survey & $A$ (m$^2$) & $D$ (m)  & $\Delta_{\text{pix}}$ & $X_{0}$  &  & (s)  & mag/arcsec$^2$ & $\kappa_0$ \\
\hline
LSST   & 32.400      &    8.360 &   $0.200''$ & 1.2 & $i$ & 5520 & 20.5    & $0.75''$  \\
       &             &          &       &   & $r$ & 5520 & 21.2    & $0.78''$  \\
\hline
HSC    &  52.810     &    8.200 &  $0.170''$ & 1.2   & $i$ & 1200 & 19.7    &
$0.56''$  \\
       &             &          &        &  & $r$ & 600  & 20.6    &
$0.67''$ \\
\hline
DES    & 10.014      &    3.934 &   $0.263''$ & 1.3 & $i$ & 900 & 20.5    & $0.79''$ \\
       &             &          &      &    & $r$ &  900 & 21.4    & $0.79''$\\
\hline
\end{tabular}
\end{center}
\caption{
Summary of the simulation parameters used to describe idealized LSST, HSC, and DES surveys: effective unobscured light-collection area $A$ in m$^2$, 
primary mirror clear aperture diameter $D$ in m, 
pixel size $\Delta_{\text{pix}}$ in arcseconds, 
median airmass $X_{0}$ as given in ~\cite{2019ApJ...873..111I, 2018PASJ...70S...4A,DESFilters},
nominal full-depth exposure time $t_{\text{exp}}$ in seconds, 
typical sky brightness $B$ in mag/arcsec$^2$, 
the full-width half-maximum (FWHM) median zenith atmospheric seeing $\kappa_0$ in arcseconds. The two numbers given in each of the last four columns are for the $i$ (upper) and $r$ (lower) bands, respectively. Sky brightness values assume a 3-day ($\sim10$\%) moon.}
\label{tab:simpar}
\end{table}

We model the atmospheric degradation of the image using a Kolmogorov point spread function (PSF) with zenith full-width half-maximum (FWHM) parameter $\kappa_0$. 
We model the instrumental PSF as an obscured Airy pattern calculated for the primary mirror diameter $D$ and obscuration by a central disk, for the central wavelength of each filter band. 
The values of the parameters used for each survey are given in \tab{simpar}.

In the following, unless otherwise stated, we assume an airmass $X = 1.2$, a round PSF
of constant size (given in \tab{simpar}), 
and no cosmic shear ($\gv = 0$). 
Our results are typically based on simulations of $13.6\times 13.6$ square arcminutes 
corresponding to one LSST 4k$\times$4k chip.

\begin{figure}
\begin{center}
\includegraphics[width=1.47in]{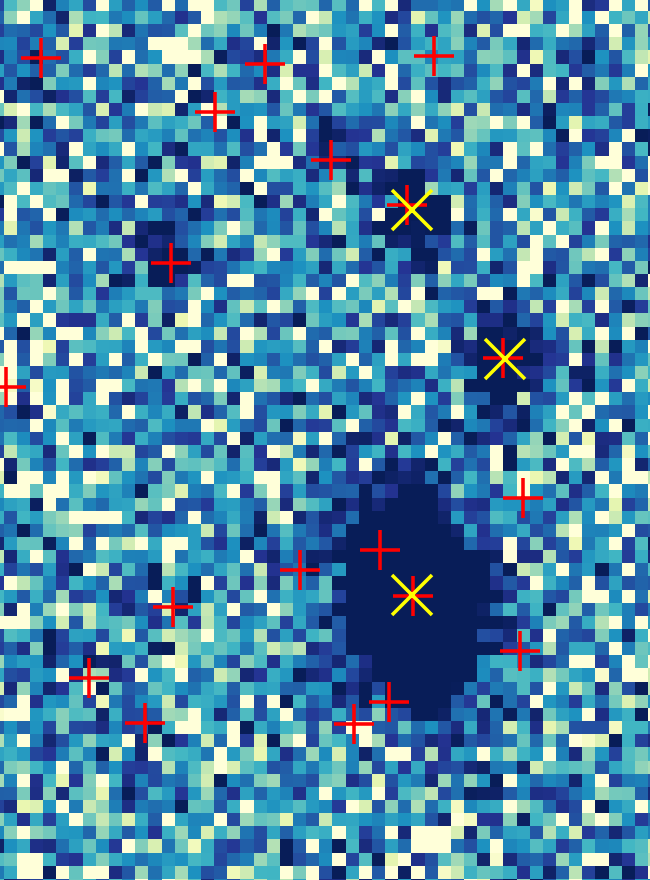}
\includegraphics[width=1.47in]{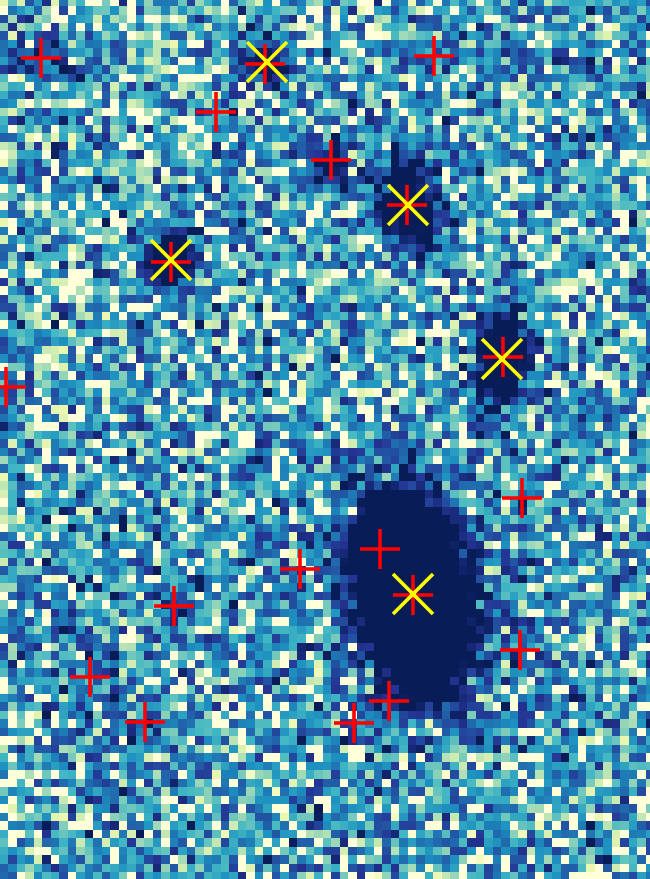}
\includegraphics[width=1.47in]{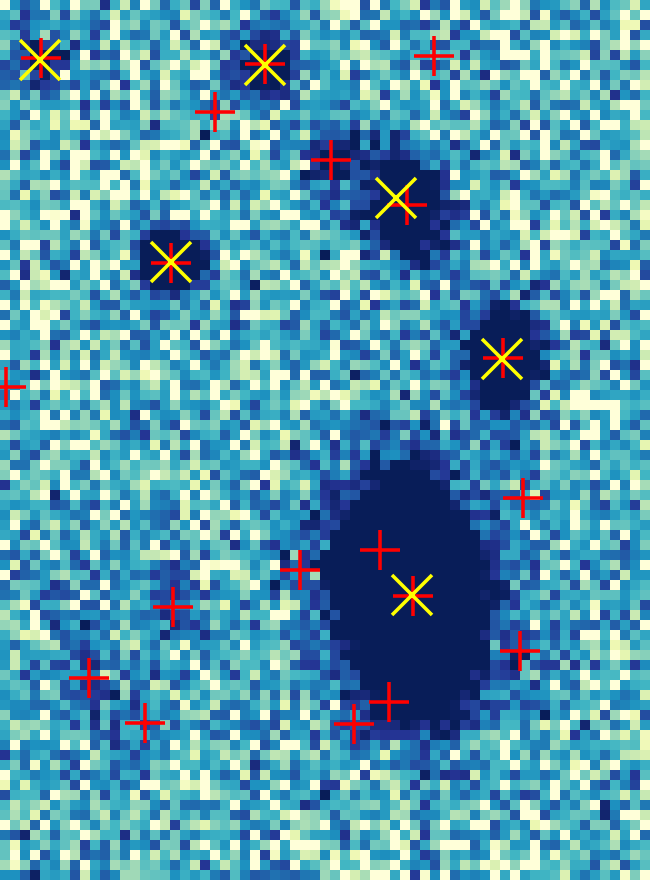}
\includegraphics[width=1.47in]{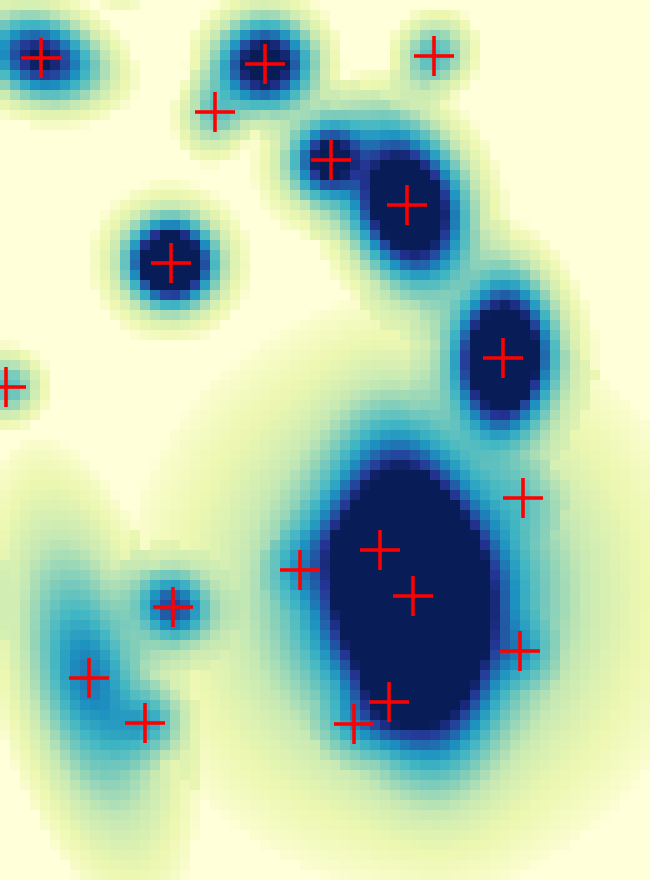}
\caption{Images of the same catalog objects simulated for (from left to right) DES $i$ band (left) and LSST $i$ band (middle and right), and covering 13.0 arcseconds by 17.6 arcseconds. The first three images include noise representative of the full survey depths given in \tab{simpar}. The right image shows noise-free LSST $i$-band predictions calculated down to a threshold $\nu_{\text{pix}} = 0.05$. The red $+$ symbols show the input centroid positions of the simulated sources. The yellow $\times$ symbols in the noisy images show the recovered centroid positions for sources detected by Source Extractor. 
}
\label{fig:sim_demo}
\end{center}
\end{figure}

Figure \ref{fig:sim_demo} shows simulated DES and LSST $i$-band images covering $13.0\times 17.6$ square arcseconds, or $65\times 88$ LSST pixels.

\section{Estimators of statistical sensitivity and pixel-noise bias based on Fisher formalism}
\label{sec:Fisherformalism}

One option for extracting galaxy shape parameters is model fitting, and one type of model-fitting algorithm is a maximum likelihood (ML) fit. 
Maximum likelihood point estimators (MLE) -- for example, for shear -- are statistically unbiased only in the limit of high signal-to-noise ratio (SNR).\footnote{Because the shear estimators are a nonlinear function of image intensity, pixel noise can lead to biased shape measurements even when the estimate of image intensity is unbiased.}  
Given a parametrized galaxy model and assuming Gaussian pixel noise, the Fisher formalism can be used to efficiently forecast the statistical sensitivity~\cite{Fisher:1935} and estimate the noise bias~\cite{Cox:1968} of MLEs for galaxy parameters relevant to shear estimation. 
Although the Fisher formalism is strictly valid only in the linearized-signal or high SNR limit~\cite{LIGO:Fisher},
it avoids the need for large numbers of simulated data sets, each with a different noise realization, for different values of SNR -- which can bring in issues of unstable fits and introduce other types of bias. 
Importantly, the Cramer-Rao theorem states that the variance of any unbiased estimator is at least as high as the inverse of the Fisher information. Therefore, the statistical sensitivities that we calculate here cannot be exceeded by any unbiased estimator. 
The results presented in this work are based on the numerical estimation of the Fisher matrices via image simulations. Galaxies and stars are rendered using the parameters from the \textsc{CatSim} catalog described in sections~\ref{sec:galaxy_catalog} and~\ref{sec:stellar_catalog}.  We  study only single-band images; images in additional bands could improve shear sensitivity, both because of the higher signal and potential improvements in deblending performance due to color information.

The simulations described in the earlier sections are used to predict the mean contribution $\mu_p$ to pixel $p$ of the final image as
\begin{equation}
\mu_{p} = b + \sum_i s_{ip} ,
\label{eqn:pixelflux}
\end{equation}
where the summation is over all sources $i$ that contribute some flux above a preset threshold ($s_{ip}\ge s_{\rm min}$ where $s_{\rm min} = 0.05 \, b^{1/2}$) to pixel $p$. In addition to calculating $s_{ip}$ for each source, we also calculate its partial derivatives $s_{ip,\alpha} = \frac{\partial{s_{ip}}}{\partial{\theta_\alpha}}$ for the following six parameters $\theta_\alpha$:
\begin{itemize}
\item Total flux $f$ in electrons that would be detected from this source in the absence of any edge cutoffs.
\item Centroid positions $x$ and $y$ in the image, measured in arcseconds.
\item Dimensionless radial scale factor $t$, applied so that flux is conserved when $t \ne 1$. 
\item Shear components $g_1$ and $g_2$ (see \eqn{lensing}). 
\end{itemize}
These partial derivatives are calculated numerically using centered first-order finite differences of postage-stamp images simulated with small variations applied to each catalog source model using the \texttt{galsim} functions \texttt{dilate} for $t$ and \texttt{shear} for $g_1$ and $g_2$. Partial-derivative images $s_{ip,\alpha}$ are truncated to the same limiting isophote as $s_{ip}$. 
Example images are shown in \fig{fisher}\footnote{ In \fig{fisher} and for the rest of the paper, HLR will refer to the unconvolved half-light radius of the galaxy being considered.} and discussed below.

For the 70\% of sources with only a single extended component, the parameter $t$ modifies only the component's half-light radius while $g_1$ and $g_2$ modify only the ellipse parameters $q$ and $\beta$. 
For the 30\% of sources with both disk and bulge components, described by an additional three parameters (relative flux, size, and ellipticity), $t$, $g_1$, and $g_2$ modify the size and ellipticity of the combined disk and bulge, since the \texttt{galsim} functions mentioned above operate on the whole profile. 

The partial derivatives for stars are calculated for only three parameters:  total flux $f$ and centroid positions $x$ and $y$.

\subsection{Statistical sensitivity estimates}
\label{sec:statsensitivity}
We use a Fisher-matrix formalism~\cite{1997ApJ...480...22T} to estimate the statistical uncertainties and noise bias on our chosen six parameters under various assumptions. The Fisher matrix ${\cal F}_{\alpha\beta}$ is
\begin{equation}
{\cal F}_{\alpha\beta} = \mu_{,\alpha}C^{-1}\mu_{,\beta} + \frac{1}{2}\text{tr}\left(
C^{-1}C_{,\alpha}C^{-1}C_{,\beta}\right)\, ,
\end{equation}
where $\alpha$, $\beta$ index the free parameters, and the pixel count $\mu$, the covariance $C$, and their partial derivatives (denoted by the subscript comma notation) are evaluated using the known true values of these parameters. We assume that fluctuations in pixel values are dominated by Poisson fluctuations in a large number of detected electrons, so that the noise in different pixels $p$ and $q$ is uncorrelated and the covariance is given by 
\begin{equation}
C_{pq} = \mu_p\,\delta_{pq} \; .
\end{equation}
With this assumption, we calculate
\begin{equation}
{\cal F}_{\alpha\beta}= \sum_{\text{pixels}~p} \left( \mu_p^{-1} + \frac{1}{2}\mu_p^{-2}\right) \, \mu_{p,\alpha} \mu_{p,\beta} \; .
\label{eqn:fisher}
\end{equation}
The quantity inside the summation over pixels gives the spatial distribution of information available on the inverse covariance of parameters $\theta_\alpha$ and $\theta_\beta$ and is useful to display as an image. 

\begin{figure}[ht]
\begin{center}
\includegraphics[width=6.0in]{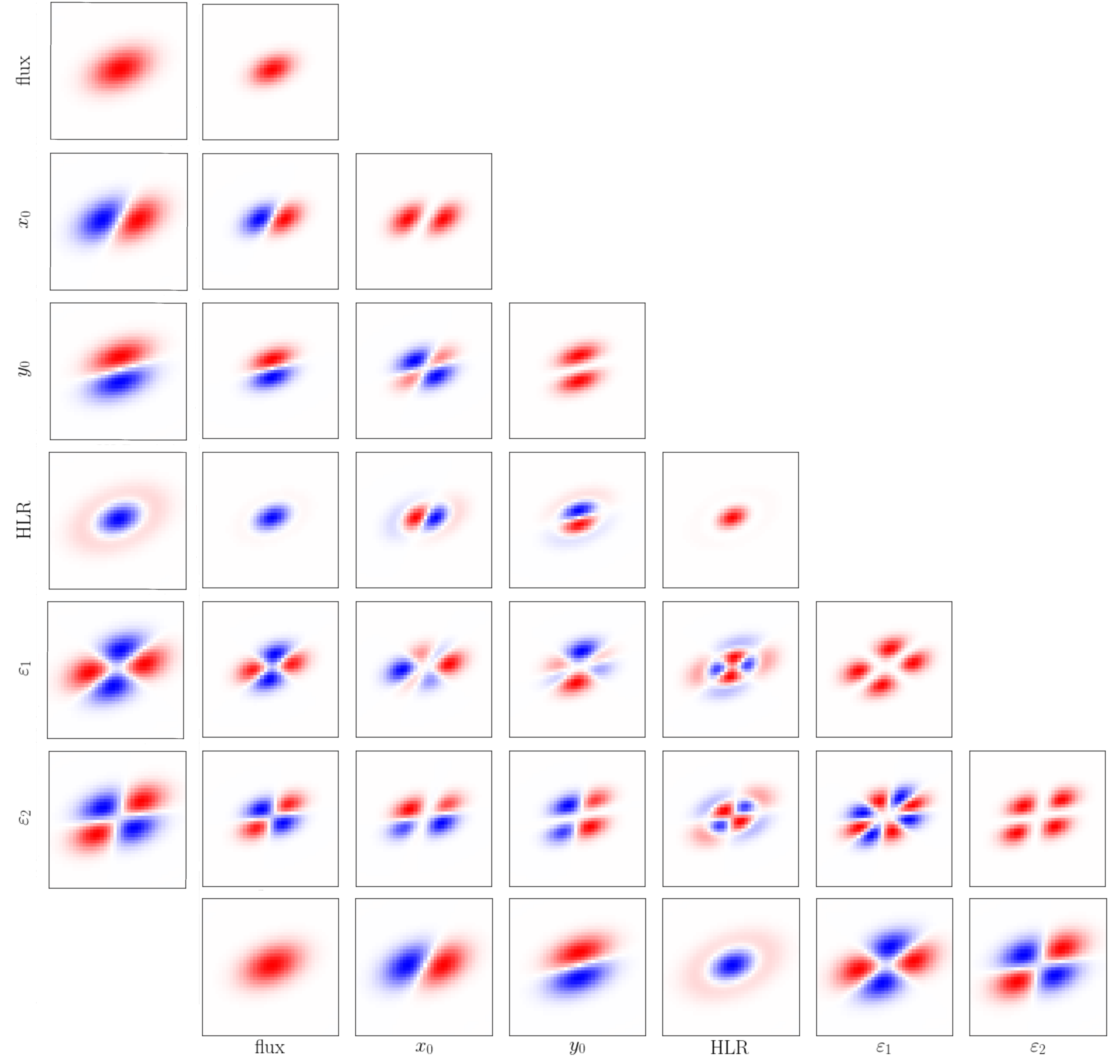}
\caption{Example of partial-derivative images (bottom row and leftmost column) and the Fisher-matrix images (middle six-by-six array, with symmetric upper-diagonal images omitted) resulting from the element-wise multiplication of the partial-derivative images to the left and bottom of each position. 
The value for pixel $p$ in row $\alpha$ and column $\beta$ corresponds to one element in the summation in \eqn{fisher}, for pixel index $p$ and parameters $\theta_\alpha$ and $\theta_\beta$.
Labels on the axes indicate the partial derivatives associated with each row and column: flux, centroid positions $x_0$ and $y_0$, HLR, $\varepsilon_{1}$, and $\varepsilon_{2}$. 
The color scale ranges from blue (negative) to red (positive), with white for zero values. 
The ranges are chosen independently for each image. 
Absolute pixel values should not be compared between images since they generally have different units, depending on the partial derivatives involved. 
The size and ellipticity components for the generated galaxy are ${\rm HLR} = 1.0$ arcsec, $\varepsilon_1 = \varepsilon_2 = + 0.2$. 
The image includes a convolution with a Gaussian PSF with $\mathrm{HLR} = 0.35$\,arcsec.
Note that the flux derivative image (lower-left corner) is a scaled version of the noise-free source image.}
\label{fig:fisher}
\end{center}
\end{figure}

Figure\,\ref{fig:fisher} shows such images for a sample galaxy. 
From the partial-derivative images in the left column and bottom row, we see that, as expected, an increase in flux in pixels to the left of the original centroid leads to a shift to the left for the centroid position $x_{0}$, and an increase in flux in pixels within the half-light radius leads to a decrease in the value of the HLR.
From the Fisher-matrix images in the remaining cells, we see, for example, that the brightest pixels near the image centroid are not informative about the centroid position or ellipticity components, but provide most of the information about the source flux and radial scale.

The factor
\begin{equation}
\mu_p^{-1} + \frac{1}{2}\mu_p^{-2}
\label{eqn:variance_norm}
\end{equation}
appearing in \eqn{fisher} determines the normalization of the estimated uncertainties due to the assumed variance in pixel values. 
 In the sky-dominated limit -- i.e., when $\mu_p \simeq b \gg 1$ -- the first term dominates and the entire expression approaches $b^{-1}$. 
In fact, we assume this sky-dominated limit when calculating pixel-level partial derivatives for the images in \fig{fisher}, and for quantities shown with explicit dependence on $\nu_{\rm sky}$ (figures \ref{fig:pixel_bias}, \ref{fig:single_gal_vary_e1}, and \ref{fig:two_gal_vary_sep}).  Here, $\nu_{\rm sky}$ is the flux signal-to-noise ratio for an optimally weighted flux estimator for faint objects, as defined in \tab{snr_defs}.\footnote{This is the same signal-to-noise ratio definition used in \textsc{GalSim}.}
In all other cases, \eqn{variance_norm} is used. 

An isolated galaxy has six unknown parameters, leading to a $6 \times 6$ Fisher matrix\footnote{The Fisher matrix is $3 \times 3$ for stars.}, but the parameters of an overlapping group of $N_{\text{grp}}$ galaxies must all be considered simultaneously, leading to a $6 N_{\text{grp}} \times 6 N_{\text{grp}}$ Fisher matrix.  In our analysis of blended sources, we calculate two sets of estimated uncertainties: first, treating each source as if it were isolated and ignoring correlations with other source parameters, and, second, including the full set of correlations.  Given a set of correlated parameters, there are two ways to estimate the individual uncertainties on each parameter $\theta_\alpha$. We can either assume that the values of all other parameters are known, leading to
\begin{equation}
\tilde{\sigma}_{\alpha} \equiv {\cal F}_{\alpha\alpha}^{-1/2} \; ,
\label{eqn:sigma1}
\end{equation}
or else marginalize over the unknown values of all other parameters, leading to
\begin{equation}
\sigma_{\alpha} \equiv ({\cal F}^{-1})_{\alpha\alpha}^{+1/2} \; .
\label{eqn:sigma2}
\end{equation}
The different uncertainty estimation models we use in the following are summarized in \sect{snr_purity} with the flux parameter, $\theta_\alpha = f$, used as an example. Note that since the signal and sky background are both linear in the exposure time $t_{\text{exp}}$, the same is true of $\mu_p$ and its partial derivatives; therefore all estimated uncertainties scale with $t_{\text{exp}}^{-1/2}$ in the usual limit that $\mu_p \gg 1$.

To reduce sensitivity to numerical precision issues during inversion, we apply an equilibration procedure to precondition the Fisher matrix by rescaling the flux parameter. 
If the Fisher matrix after equilibration is not invertible or any of the variances in $C$ are less than or equal to zero, we drop the source with the lowest value of signal-to-noise ratio $\nu_{\rm iso}$ (defined in sec.\,\ref{sec:snr_purity}) and attempt to invert again. 
This procedure is iterated until we have a valid covariance matrix. Fisher matrix elements for all parameters of any sources that are dropped in this procedure are set to zero and variances are set such that $1/\sqrt{\sigma_{\alpha}^{2}}=0$. 
Invalid covariances are generally associated with sources that are barely above the pixel threshold, which is to be expected since a six-parameter galaxy model cannot be constrained unless at least six pixels are above threshold; therefore, this procedure should normally provide sensible values for the largest possible subset of input sources. 

\subsection{Noise bias formalism} \label{sec:noisebias}
\begin{figure}[tbp]
\includegraphics[width=\textwidth]{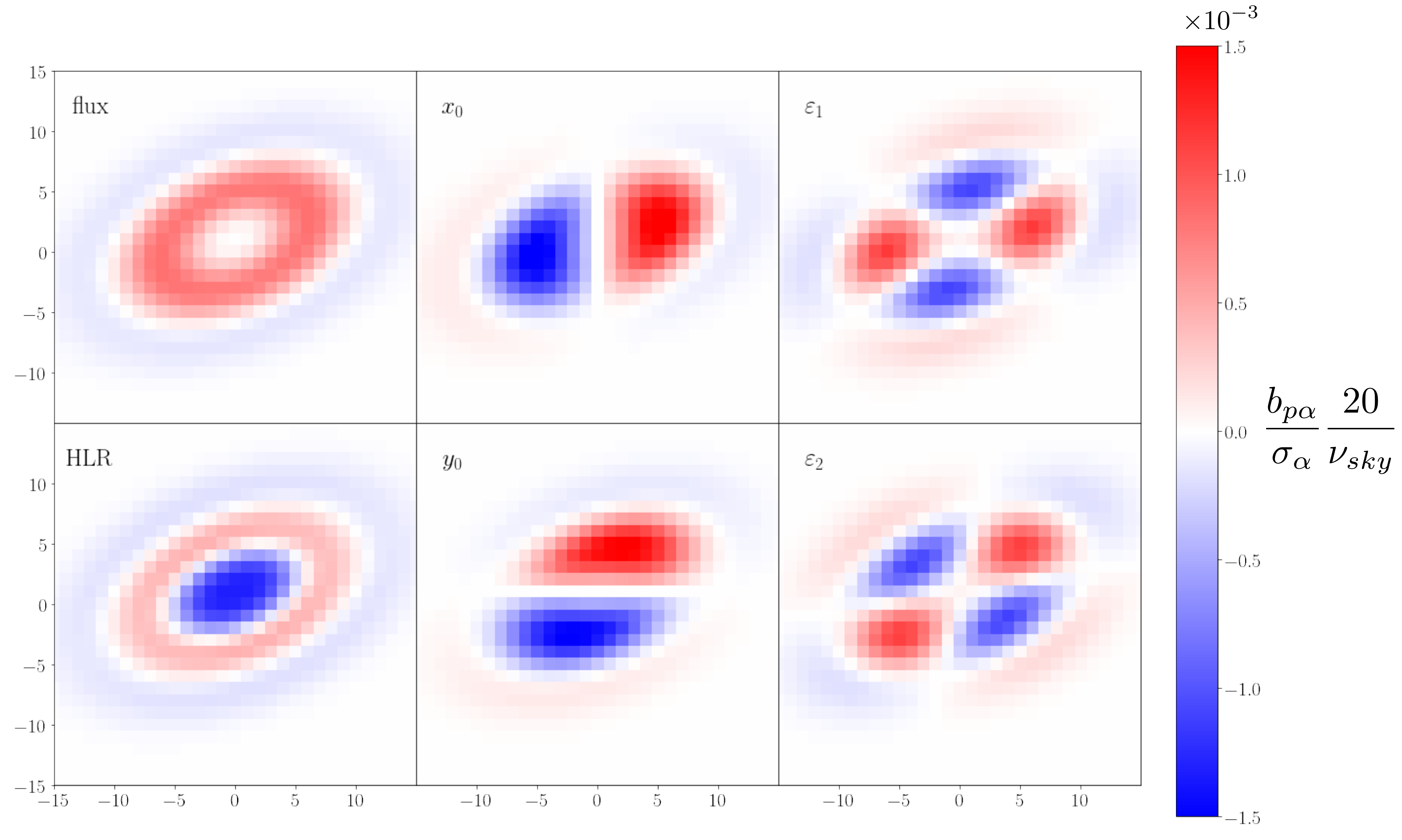}
\centering
\caption{Images of $(b_{p\alpha}/\sigma_{\alpha}) (20/\nu_{\rm sky})$, where $b_{p\alpha}$ is the bias contribution per LSST pixel $p$ for parameter $\theta_\alpha$, and $\sigma_{\alpha}$ is the Fisher prediction for the uncertainty, calculated for the same sample galaxy used to generate \fig{fisher}. We include the factor of $(20/\nu_{\rm sky})$ to make explicit the Fisher prediction for the scaling with SNR.
The axes on the images correspond to LSST pixels.}
\label{fig:pixel_bias}
\end{figure}

The bias ${b}_{\alpha}$ for parameter $\theta_\alpha$ is defined as   
  \begin{equation}
   b_{\alpha} \equiv \langle \hat{\theta}_\alpha - \theta_\alpha \rangle\; ,
  \label{eqn:bias_def}
  \end{equation}
where $\theta_\alpha$ is the true value of the parameter, $\hat{\theta}_\alpha$ is the ML estimator, and the expectation value is over many noise realizations. 
The Fisher formalism can be used to show that the expected noise bias is given by the following expression \cite{Cox:1968}:
  \begin{equation}
  b_{\alpha} =  - \frac{1}{2} \, \mathcal{C}_{\alpha\beta} \, \mathcal{C}_{\gamma\delta}\, \mathcal{B_{\beta\gamma\delta}} + \mathcal{O}\left(\nu^{-4}\right)\; ,
  \label{eqn:bias}
  \end{equation}
where the $\mathcal{C}_{\alpha\beta}$ are the elements of the covariance matrix given by the inverse of the Fisher matrix in \eqn{fisher} and $\nu$ is the SNR. 
(Summation over repeated Greek-letter indices is assumed here and thereafter.)
$\mathcal{B_{\beta\gamma\delta}}$ is given by
  \begin{equation}
  \mathcal{B}_{\beta\gamma\delta}= \sum_{\textnormal{\scriptsize pixels}~p} \left( \mu_p^{-1} + \frac{1}{2}\mu_p^{-2}\right) \, \mu_{p,\beta} \, \mu_{p,\gamma\delta} \; , 
  \label{eqn:bias_tensor}
  \end{equation}
where $\mu_p$ is the pixel flux defined in \eqn{pixelflux}, 
$\mu_{p,\alpha}$ is its partial derivative with respect to $\theta_\alpha$, 
and $\mu_{p,\gamma\delta}$ is its second partial derivative with respect to $\theta_\gamma$ and $\theta_\delta$.
The expected bias can then be written as 
  \begin{equation}
  b_{\alpha} = \sum_{\textnormal{\scriptsize pixels}~p}\, \left( - \frac{1}{2}\right) \left( \mu_p^{-1} + \frac{1}{2}\mu_p^{-2}\right)\, \mathcal{C}_{\alpha\beta} \, \mathcal{C}_{\gamma\delta}\, \mu_{p,\beta} \, \mu_{p,\gamma\delta} \; + \mathcal{O}\left(\nu^{-4}\right) \; ,
  \label{eqn:full_bias}
  \end{equation}
where we used the fact that the covariance matrix elements do not have any pixel dependence to bring them into the sum over pixels. 
We can then define the bias contribution $b_{p\alpha}$ per pixel as  
  \begin{equation}
  b_{p\alpha}=  \left( - \frac{1}{2}\right) \left( \mu_p^{-1} + \frac{1}{2}\mu_p^{-2}\right) \, \mathcal{C}_{\alpha\beta} \, \mathcal{C}_{\gamma\delta}\, \mu_{p,\beta} \, \mu_{p,\gamma\delta} \; ,
  \label{eqn:biasperpixel}
  \end{equation}  
and the total bias as
  \begin{equation}
  b_{\alpha} = \sum_{\textnormal{\scriptsize pixels}~p}b_{p\alpha} \; .
  \label{eqn:fisher_bias}
  \end{equation}
In \fig{pixel_bias}, we show images of the pixel bias $b_{p\alpha}$ given in \eqn{biasperpixel}, divided by the Fisher prediction for the uncertainty $\sigma_{\alpha}$ (as defined in \eqn{sigma2}) for parameter $\theta_\alpha$, for the same original sample image illustrated in \fig{fisher}. This figure shows how different parts of a galaxy image can have a positive or negative contribution to the bias of the corresponding maximum-likelihood estimator. For example, pixels close to the center of a galaxy contribute significant negative bias to the HLR estimator, but these pixels do not contribute significant bias to the flux estimator.

The factor of $(20/\nu_{\rm sky})$ makes explicit the Fisher prediction for the scaling of noise bias and uncertainty with SNR in the sky-dominated limit.
The pixel size in the bias image corresponds to the LSST pixel size.
Hence, the numerical scale for each image corresponds to the bias contribution per LSST pixel, expressed as a fraction of the expected uncertainty $\sigma_{\alpha}$, with red and blue corresponding to positive and negative bias, respectively.

Given the symmetry of the image about the centroid, we do not expect a net bias for the centroid position $(x_0,y_0)$.  However, a nonzero bias is possible for the other four parameters for this non-circular image.

\section{Validation tests for isolated galaxies}
 
In this paper, we use the Fisher-matrix formalism to characterize  parameter-estimation performance of MLE for galaxy shape measurements -- for both the covariance matrix and noise bias -- given parameterized models of the image profiles, and assuming known Gaussian pixel noise. 
The Fisher matrix can be a poor predictor of information content when there are a significant number of (possibly strongly correlated) parameters and the expected SNR is relatively low (or, equivalently, when the assumption that the profile depends linearly on the parameters is no longer valid), or when the profile depends weakly on one or more parameters and their priors are not taken into consideration; see ref.~\cite{LIGO:Fisher}, for example, for a discussion of pitfalls. 
Therefore, we investigate the range of validity of the Fisher formalism and then carefully choose a galaxy sample for the analysis that minimizes these pitfalls, as described in  \sect{sample_definitions}. 

In \app{dep_on_size_ell}, we illustrate (for isolated galaxies) how noise bias on galaxy size, shape, and ellipticity parameters depends on the galaxy shape itself. 
We show that the sign and magnitude of the predicted bias on the shear estimator depends sensitively on which shape estimator is used and the relative size of the galaxy and the PSF.

We used ML fits to a simulated sample of 40,000 noise realizations of a single galaxy to show numerically that the Fisher estimates of the covariance matrix and noise bias for all parameters are accurate within the statistical sensitivity provided by the sample. 

We find that our results are generally consistent with prior studies of estimates of pixel-noise bias for isolated galaxies \cite{Hirata&S2004, Melchior:2012un, Refregier:2012kg, Kacprzak:2012kf, HallTaylor2017, OkuraFutamase2018}. 
\section{Definitions of source signal-to-noise ratio and purity}
\label{sec:snr_purity}
In parts of this study, we classify sources as ``detectable'' or not based on the value of a particular definition of signal-to-noise ratio (SNR).   
We further classify detectable sources as ``blended'' or not based on the source's ``purity''  
-- i.e., the degree to which it overlaps with other sources. 
In this section, we define these measures of SNR and purity used to classify sources. 

The criterion we use to classify a source as ``detectable'' is whether its appropriately defined signal-to-noise ratio $\nu$ is above some detection threshold. We generically calculate signal-to-noise ratios as
\begin{equation}
\nu = \frac{f}{\sigma_f} \; ,
\label{eqn:nu_gen}
\end{equation}
where $f$ is the unknown source flux  
and $\sigma_f$ is one of the flux uncertainty estimates defined in \tab{snr_defs}.

\begin{table}[tbp]
\begin{center}
\begin{tabular}{lcll}
Assumptions & Subscript & Assumed variance & Free parameters \\
\hline
Sky dominated & sky & $b$ & $f_i$  \\
Isolated, free & isof & $b + s_{ip}  + \ldots$ & $f_i,x_i,\ldots$  \\
Group, free & grpf & $b + s_{ip} + s_{jp} + \ldots$ & $f_i,x_i,\ldots,f_j,x_j,\ldots$\\
\hline
\end{tabular}
\caption{
Definitions of the different Fisher-matrix uncertainty-estimation models used to define flux uncertainty $\sigma_f$ and SNR estimates $\nu = f/\sigma_f$ in our analysis. The third column lists the variance that is used to compute the Fisher matrix, using \eqn{variance_norm} (except for the ``sky'' case, where \eqn{variance_norm} is replaced with $b^{-1}$). The fourth column lists the parameters that are considered free when marginalizing the Fisher matrix. In both columns, the index $i$ denotes the galaxy whose quantities are being estimated and the index $j$ denotes any other overlapping galaxies.}
\label{tab:snr_defs}
\end{center}
\end{table}

In the designations of SNR $\nu$, the subscript ``isof'' indicates that each object is simulated in a separate image (blending off), and ``grpf'' indicates that an entire group of overlapping objects is simulated in one image (blending on). 
For the first SNR estimate in \tab{snr_defs} ($\nu_{\text{sky}}$)
we effectively assume that the source's size and shape are perfectly known for the purposes of flux estimation -- i.e., $\sigma_f = ({\cal F}_{ff})^{-1/2}$ from \eqn{sigma1}. 
For the final two estimates ($\nu_{\text{isof}}$,  $\nu_{\text{grpf}}$), we marginalize over the free parameters listed in the third column so that $\sigma_f = ({\cal F}^{-1})_{ff}^{+1/2}$ from \eqn{sigma2},  
leading to more realistic signal-to-noise ratios.

Galaxies may be detectable yet sufficiently blended that an accurate and unbiased measurement of their shear response is challenging.  We define a source's ``purity'' $\rho$ as a measure of the degree of blending:
\begin{equation}
\rho_i \equiv \frac{\sum_p s_{ip}\cdot s_{ip}}{\sum_p (s_{ip}\cdot \sum_j s_{jp})} \; ,
\label{eqn:rho}
\end{equation}
where the sums over $p$ are over all pixels within the overlapping group and the sum over $j$ is over all sources with any overlap with source $i$, including $i$ itself. 
Purity is then a ratio of weighted flux estimates over pixels, where we treat the object as being isolated in the numerator and include overlaps in the denominator, and use the true profile of source $i$ for the weights in both cases. 
By construction, $0 < \rho_i \le 1$ with $\rho_i = 1$ for completely isolated sources. 

We set a threshold of $\rho = 0.98$ to divide the sample of all simulated galaxies
into similar numbers of low-purity and high-purity sources.

Based on comparisons with Source Extractor (SE)~\citep{1996A&AS..117..393B} detections\footnote{We use version 2.19.5 of Source Extractor with the settings specified in \app{sextractor_settings}.}, we somewhat arbitrarily define a galaxy (either high-purity or low-purity) to be ``detectable'' if its value of $\nu_{\text{grpf}}$ is above a threshold that corresponds to a detection rate of approximately 50\%. 
In an LSST full-depth $i$-band observation this corresponds to a $\nu_{\text{grpf}}$ threshold close to 6.  
For both DES and HSC full-depth $i$-band observations, a detection rate of 50\% corresponds to a $\nu_{\text{grpf}}$ threshold close to 5. The higher threshold for LSST is due to the greater depth compared to DES or HSC (and larger PSF than HSC) and subsequently higher fraction of blended objects, making detection more challenging for Source Extractor; this leads to a higher $\nu_{\text{grpf}}$ threshold for LSST compared to DES or HSC, to accept the same fraction of sources.

We find that for approximately 62\% of sources with $\nu_{\rm grpf} > 6$ in full-depth LSST images, at least 1\% of the flux in their pixels is due to overlapping sources.

\begin{figure}[tbp]
\begin{center}
\includegraphics[width=6.0in]{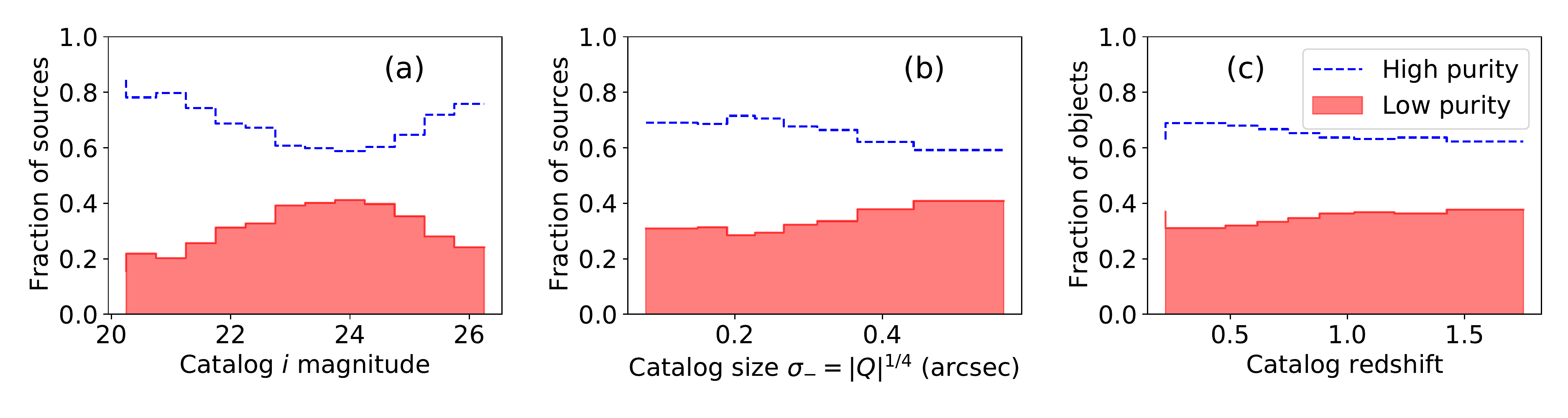}
\caption{Binned fraction of detectable galaxies ($\nu_{\rm{grpf}} > 6$) that are high-purity ($\rho \geq 0.98$, blue dashed line) and low-purity ($\rho < 0.98$, red filled histogram) as a function of three catalog quantities: galaxy magnitude, size, and redshift. The plots are based on a simulated LSST full-depth single-sensor $i$-band exposure.}
\label{fig:samples}
\end{center}
\end{figure}

In \fig{samples}, we plot the binned fraction of detectable galaxies ($\nu_{\rm{grpf}} > 6$) that are high-purity ($\rho \geq 0.98$) and low-purity ($\rho < 0.98$) as a function of three quantities: galaxy magnitude, size, and redshift. 
We observe a larger ratio of high-purity to low-purity galaxies for both the brightest and faintest galaxies (panel a). This is expected for the brightest galaxies since they will also have higher purity, on average. For the faintest galaxies, the explanation is that fainter galaxies that are also low purity will not satisfy the minimum SNR criteria for detection. 
Similar logic explains the dependence of the fraction of low-purity objects on size (panel b). Smaller galaxies will typically be fainter; hence, small low-purity objects are less likely to satisfy the minimum SNR for detection than small high-purity (brighter) objects. 
We observe a mild redshift dependence (panel c) due to low-redshift galaxies being brighter (and higher purity) on average.

\section{Results: Loss in statistical sensitivity due to blending}

We analyze the simulated images described above to quantify two impacts of overlapping sources on our ability to estimate cosmic shear: loss of statistical sensitivity and increase in pixel-noise bias.
In this section, we review the formalism used to forecast loss of sensitivity and present the results first for galaxies alone, and then in the presence of stars.
The impact of blending on pixel-noise bias is presented in sec.\,\ref{sec:Pixel-noise bias}. 


\subsection{Definition of effective number density, \texorpdfstring{$N_{\rm eff}$}{Neff}}
\label{sec:Neff}

For the purposes of shear estimation, the shape of a galaxy is described by a two-component spinor $\epsv$, so its statistical uncertainty is described by the $2\times 2$ submatrix of a larger covariance matrix where correlations with other parameters have been marginalized over.  With $N_{\text{grp}}$ overlapping galaxies, the full covariance matrix includes correlations between the spinor components of different galaxies, leading to a $2N_{\text{grp}} \times 2N_{\text{grp}}$ submatrix of shape covariances.

We assume that intrinsic (denoted by I) unlensed shapes $\epsv_i$ of galaxies $i$ are randomly distributed and uncorrelated between galaxies, with a truncated Gaussian probability density in the two spinor components described by the $2\times 2$ covariance matrix $C_{\text{I}}$, which we take to be
\begin{equation}
C_{\text{I}} = \begin{pmatrix}
\sigma_{\text{I}}^2 & 0 \\
0 & \sigma_{\text{I}}^2 \\
\end{pmatrix} \; ,
\end{equation}
where $\sigma_{\text{I}}^2$ is the intrinsic shape-noise variance. 

We further assume that $\hat{\varepsilon}_i'$ is a noisy estimator of the true observed shape $\varepsilon_i'$ under the action of a constant shear $\gv$, \eqn{lensing}, with a $2\times 2$ covariance $C_{ij}$ between the spinor components of $\hat{\epsv}_i'$ and $\hat{\epsv}_j'$ ($C_{ij}$ is zero unless galaxies $i$ and $j$ are overlapping). The optimal linear estimator of \eqn{estimator} is then
\begin{equation}
\hat{\gv} = C_g \sum_{ij} \left(C_{\text{I}}\delta_{ij} + C_{ij}\right)^{-1} \hat{\epsv}_j' \; ,
\end{equation}
with $2\times 2$ inverse covariance
\begin{equation}
C_g^{-1} = \sum_{ij} \left(C_{\text{I}}\delta_{ij} + C_{ij}\right)^{-1} \; .
\end{equation}

In the limit of noise-free measurements ($C_{ij} \rightarrow 0$), the shear estimate uncertainty is simply due to averaging out the intrinsic covariance of $N$ galaxies,
\begin{equation}
C_g \rightarrow N^{-1}\, C_{\text{I}} \; .
\end{equation}
It is useful to assign a weight to individual high-purity or low-purity galaxies that quantifies its contribution to the overall shear estimate. There is some arbitrariness to this choice, especially for low-purity galaxies where a per-galaxy weight is not well defined; 
we adopt the definition
\begin{equation}
\omega_i \equiv \frac{1}{2} \left( \frac{\sigma_{\text{I}}^2}{\sigma_{\text{I}}^2 + \sigma_{i,1}^2} +
\frac{\sigma_{\text{I}}^2}{\sigma_{\text{I}}^2 + \sigma_{i,2}^2} \right) \; ,
\label{eqn:weights}
\end{equation}
where $\sigma_{i,k}$ is the uncertainty on measuring spinor component $k$ of $\hat{\epsv}_i$, estimated using the Fisher matrix formalism described above with the ``grpf'' parameterization of \tab{snr_defs}. With this choice, we have $0 < \omega_i < 1$ with $\omega_i \rightarrow 1$ in the limit of vanishing measurement uncertainties $C_{ii}$, and we recover the usual definition of the effective number density of perfectly measured shapes~\cite{2013PhR...530...87W,2013MNRAS.434.2121C}, $\Neffgrpf$, by summing weights over a unit area:
\begin{equation}
\Neffgrpf \equiv  \frac{\sum_i \omega_i}{ A}\, ,
\label{eqn:Neff_weights}
\end{equation}
where the sum is over all galaxies in area $A$. Note that our choice of weights is equivalent to
\begin{equation}
    \omega_i = \frac{\sigma_{\text{I}}^2}{\sigma_{\text{I}}^2 + \sigma_{i}^2}
\end{equation}
with the combined effective measurement error
\begin{equation}
    \sigma_i^2 \equiv \frac
    {\sigma_{\text{I}}^2 \left(\sigma_{i,1}^2+\sigma_{i,2}^2\right) + 2 \sigma_{i,1}^2 \sigma_{i,2}^2}
    {\sigma_{i,1}^2+\sigma_{i,2}^2 + 2 \sigma_\text{I}^2} \; .
    \label{eqn:effective_weight}
\end{equation} 

In practice, we fix the intrinsic shape noise $\sigma_I$ by averaging the variances of the components of $\hat{\epsv}$ over all detected (high-purity or  low-purity) galaxies from some assumed survey configuration.

\subsection{Predicted values of \texorpdfstring{$N_{\rm eff}$}{Neff} for LSST, HSC, and DES}
\label{sec:summary_predicted_Neff}

In~\tab{summary}, we list the intrinsic shape noise $\sigma_I$ and the predicted values of  effective number density for LSST, HSC, and DES full-depth observations in $i$ and $r$ bands. We also give the weighted mean redshift for each survey: 
$\langle z \rangle = \sum_i w_i z_i$.
We list the values of two measures of the predicted statistical shear-estimation sensitivity achievable with different survey configurations:
$\Neffgrpf$ and $\Neffpur$, where $\Neffgrpf$ corresponds to the sum of grpf weights $w_i$ for all detectable galaxies and $\Neffpur$ is the sum of grpf weights for only high-purity ($\rho > 0.98$) galaxies, in a unit area.
$\Neffgrpf$ represents the highest possible value for detectable galaxies, taking into account the correlations between blended galaxies, but is unlikely to be achievable due to low-purity galaxies.
In the second last column, we give  the ratio $\Neffpur / \Neffgrpf$ for each survey and band, which in turn allows us estimate the relative contribution of high- and low-purity sources to the total signal.
$\Neffpur / \Neffgrpf$ is $\sim 0.74$ for HSC and DES $i$ and $r$ bands, and $\sim 0.62$ for LSST $i$ and $r$ bands, quantifying the expectation that blending has a greater impact at LSST depths. Additionally, we give the ratio $\Neffgrpf/\Neffisof$, where $\Neffisof$ corresponds to the sum of isof weights for galaxies with $\nu_{\rm{isof}} > 6$. This ratio gives us an idea of the total signal lost due to overlaps. 

\begin{table}[tbp]
\begin{center}

\begin{tabular}{|lccccccc|}
\hline
Survey & band & $\langle z\rangle$ & $\sigma_{\text{I}}$ & $\Neffgrpf$ [arcmin$^{-2}$]  & $\Neffpur$ [arcmin$^{-2}$] & $\Neffpur / \Neffgrpf$ & $\Neffgrpf/\Neffisof$\\
\hline
LSST & $i$ & 1.05 & 0.245 & 37.8 & 24.0 & 0.63 & 0.83 \\
     & $r$ & 1.18 & 0.243 & 39.4 & 24.0 & 0.61 & 0.81 \\
\hline
HSC  & $i$ & 1.00 & 0.251 &  31.7 &  23.6 & 0.74 & 0.89\\
     & $r$ & 1.07 & 0.253 &  23.2 &  17.3 & 0.74 & 0.86\\
\hline
DES  & $i$ & 0.85 & 0.260 &  11.4 &  8.4 & 0.74 & 0.80\\
     & $r$ & 0.92 & 0.259 &  10.3 &  7.5 & 0.73 & 0.78\\
\hline

\end{tabular}
\caption{Summary statistics for LSST, HSC, and DES full-depth observations in the $i$ and $r$ bands. Mean redshifts are weighted averages using the weights in \eqn{weights}. $\sigma_{\text{I}}$ is the standard deviation of the intrinsic galaxy shape noise for detectable sources ($\nu_{\text{grpf}} > 6$). The value of $\sigma_{\text{I}}$ is slightly different for the two bands in each survey because the detectable sample depends on the value of $\nu_{\text{grpf}}$ for each galaxy, which is band-dependent. $\Neffgrpf$ corresponds to the sum of grpf weights $w_i$ for all galaxies with $\nu_{\rm {grpf}} > 6$, while $\Neffpur$ is the contribution to $\Neffgrpf$ due to only high-purity galaxies ($\rho \ge 0.98$). $\Neffisof$ corresponds to the sum of isof weights, $w_{i}$ for all galaxies with $\nu_{\rm{isof}} > 6$.}
\label{tab:summary}
\end{center}
\end{table}  

\begin{figure}[tbp]
\begin{center}
\includegraphics[width=6.0in]{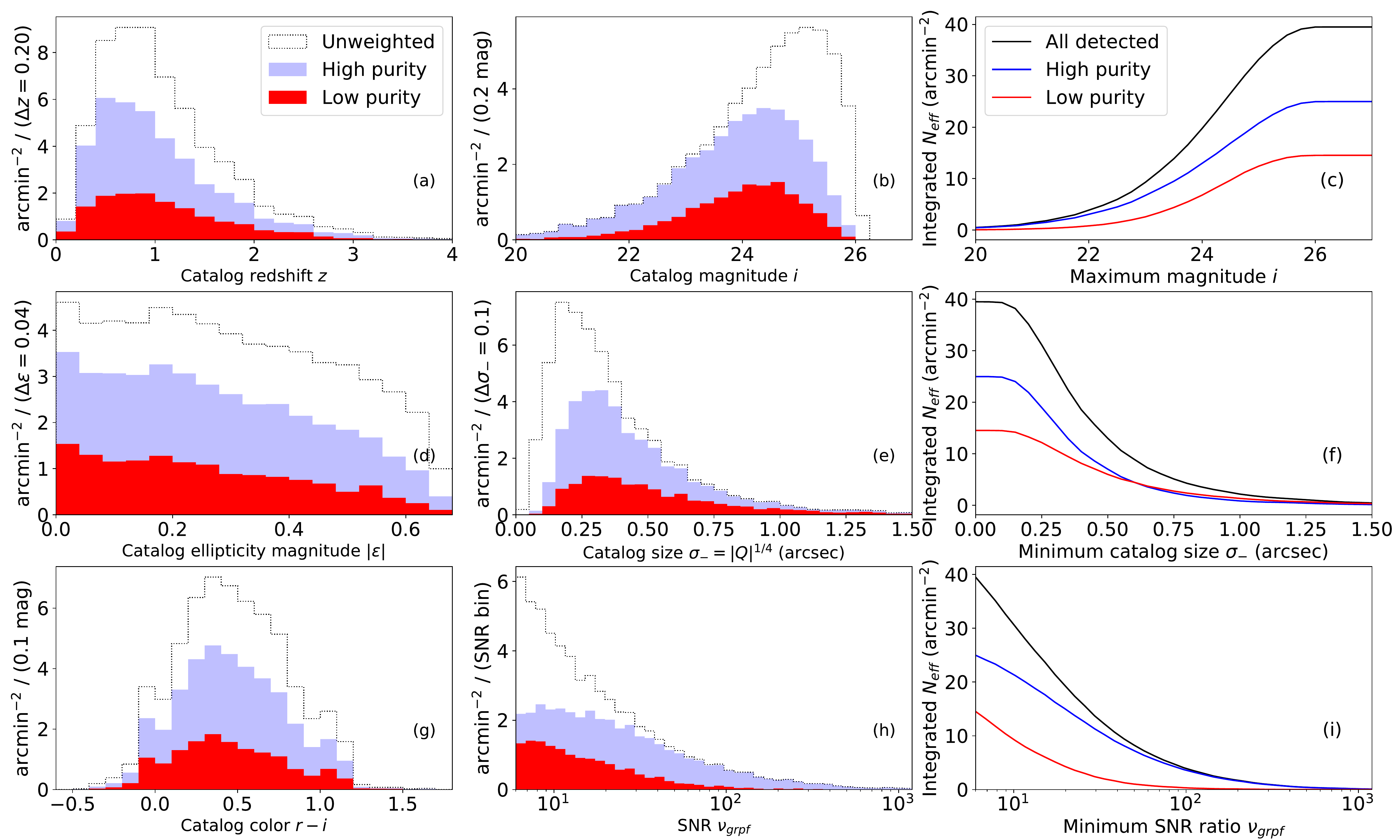}
\caption{Distributions of catalog parameters for detectable galaxies ($\nu_{\text{grpf}} > 6$) in a simulated LSST full-depth $i$-band exposure. Dotted-outline histograms show unweighted distributions while the filled red and blue regions show the stacked weighted distributions of low-purity and high-purity galaxies, respectively, using the weights $w_i$ defined in \eqn{weights}. 
The right-hand plots show the cumulative $\Neffgrpf$ as a function of (c) maximum magnitude in the simulated $i$-band filter, (f) minimum galaxy size $\sigma_-$, and (i) minimum signal-to-noise estimate $\nu_{\text{grpf}}$. }
\label{fig:LSST_i}
\end{center}
\end{figure}

In \fig{LSST_i}, for LSST $i$ band (and figures \ref{fig:HSC_i}-\ref{fig:DES_i} in \app{Neff_plots} for HSC and DES), the histograms in the first two columns show the distributions of six catalog parameters (redshift, $i$-band magnitude, ellipticity, size, color, $\nu_{\rm grpf}$) for detectable galaxies ($\nu_{\rm grpf} > 6$ for LSST and $\nu_{\rm grpf} > 5$ for HSC and DES).  The dashed histograms correspond to the \textit{unweighted} distributions; the filled red and blue regions show the stacked \textit{weighted} contributions of low-purity and high-purity galaxies, respectively, using the weights $w_i$ defined in \eqn{weights}. 
In other words, the difference between the dashed histogram and the top of the blue regions is due to the loss in statistical sensitivity due to  measurement uncertainties.
The plots in the right-hand column of each figure show the integrated $\Neffgrpf$ (i.e., weighted number density) as a function of three parameters -- maximum AB magnitude in the simulated filter, minimum galaxy size $\sigma_-$, and minimum signal-to-noise estimate $\nu_{\text{grpf}}$ -- for all detectable galaxies, and for those with high and low purity. These plots allow us to identify the attributes of the galaxies that contribute most and least significantly to  $\Neffgrpf$. For example, despite there being a non-negligible fraction of objects with $\sigma_{-} < 0.25$, or with $i$-band magnitude greater than 25.5 (see~\fig{catalog_plots}), we observe in panels (c) and (f) that these objects do not add significant information in a cosmic shear analysis. This is expected since shape measurements of small faint objects will have limited statistical significance. A similar behavior is expected for low SNR galaxies;  however, we do not see this in panel (i) due to the SNR requirement ($\nu_{\text{grpf}}>6$) for detectable galaxies, applied here.

We see that for LSST, the value of $\Neffgrpf$ in $r$ band (39.4  galaxies per arcmin$^2$) is slightly higher than in $i$ band since a higher signal-to-noise is expected for more objects in $r$ band. However, the contribution of the low-purity objects is similar in both cases (between 37\% and 39\%). For DES we see that $\Neffgrpf$ saturates in the $i$ band around $i=24$, as expected for this survey, and that the contribution of low-purity objects to $\Neffgrpf$ is much lower (26\% for $r$ band and 27\% for $i$ band). For HSC, $\Neffgrpf$ saturates at $i=25$ and $r= 25$. Since the signal-to-noise is better in $i$ band for this survey, a higher integrated $\Neffgrpf$ is recovered in $i$ band. The contribution of low-purity objects is similar for both bands (approximately 26\% for both $r$ band and $i$ band). 

In the three surveys,  the results for $\Neffgrpf$ suggest that going fainter than $r=24$ implies a non-negligible contribution of low-purity sources to the weak-lensing signal. Thus, if we think of the low-purity sources as the ones most affected by blending, studying and improving deblending techniques, and further developing techniques such as those presented in refs.~\cite{2020ApJ...902..138S, 2020arXiv201208567M}, will be critical for forthcoming galaxy surveys such as LSST in order to achieve the desired level of statistical sensitivity. 

We also use the Fisher formalism to compute $\Neffgrpf$ in the sky-dominated limit ($\mu_{p} \sim b)$. We find that in this limit  $\Neffgrpf$ is only 0.2\% larger than the values in \tab{summary}.
\subsection{Comparison with prior studies of \texorpdfstring{$N_{\rm eff}$}{Neff}}
\label{sec:prior_studies}

Earlier studies using simulated LSST images~\cite{2013MNRAS.434.2121C}\footnote{This earlier study also used the \textsc{CatSim} catalog.}, or real data from DES~\cite{2016MNRAS.460.2245J} and HSC~\cite{2018PASJ...70S..25M}, provide estimates for $N_{\rm eff}$ that can be compared with our results and, in the case of DES and HSC, used to check the realism of our predictions.
To make this comparison, we estimate $\Neffgrpf$ using galaxy samples selected with criteria designed to emulate those applied in the aforementioned studies and mention some limitations to this approach. In this section, we summarize the comparisons with previous studies, and refer the reader to~\app{prior_studies} for details. We find that our estimates of $\Neffgrpf=5.8$ galaxies/arcmin$^{2}$ for DES and $\Neffgrpf=22.3$ galaxies/arcmin$^{2}$ for HSC are very close to those found in the literature: $\Neffgrpf=5.7$~\cite{2016MNRAS.460.2245J}, and $\Neffgrpf=21.8$~\cite{2018PASJ...70S..25M} galaxies/arcmin$^{2}$, respectively. For LSST we find overall larger values $\Neffgrpf$ than those found in ref.~\cite{2013MNRAS.434.2121C} for the pessimistic, fiducial, and optimistic cases showcased in that study. However, a direct comparison is somewhat difficult since, on the one hand, the spectral throughput and detector response used in this study is reduced by 15\% when compared to ref.~\cite{2013MNRAS.434.2121C}. On the other hand, we find that our predicted values for the shear measurement uncertainty $\sigma_{i}$ are typically lower than those found in ref.~\cite{2013MNRAS.434.2121C}, compensating for the loss in ``detectable" objects, and increasing the overall $\Neffgrpf$. 
\subsection{Impact of stars}
\label{sec:impact_of_stars}

In this section, we show how the presence of stars at different number densities reduces the statistical sensitivity to weak lensing. Specifically, for each of the six pointings (with different stellar densities) described earlier, we superimpose an image corresponding to the stellar catalog on the same one-square-degree galaxy image, and recalculate $\Neffgrpf$.

\begin{figure}
\centering
\includegraphics[width=0.85\textwidth]{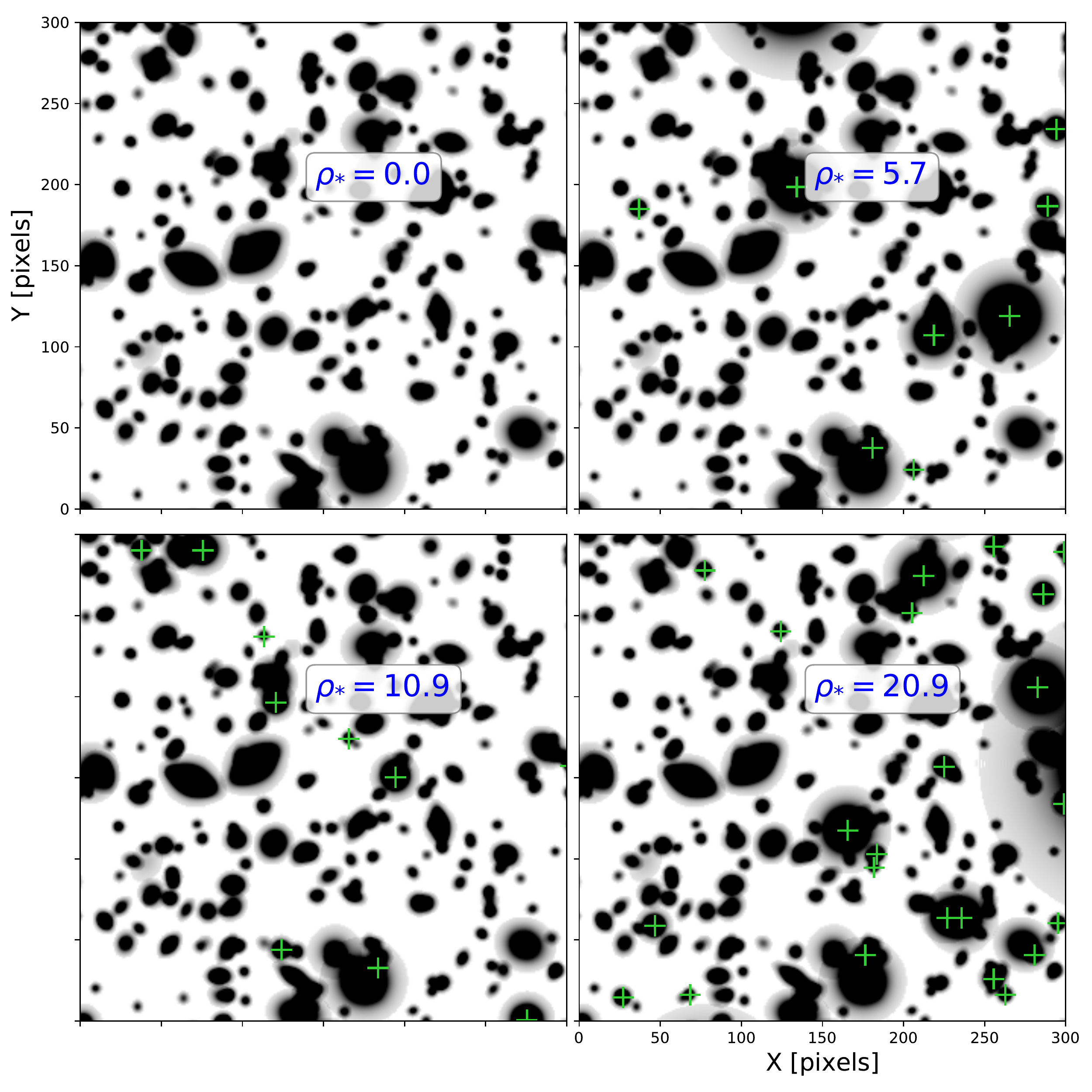}
\caption{Example of a simulated 256 pixel $\times$ 256 pixel LSST $i$-band image (0.73\,arcmin$^{2}$) with the same galaxies but stars with different stellar densities. 
The label on each cutout corresponds to the value of the added stellar density $\rho_*$ in units of stars arcmin$^{-2}$. 
Each star is indicated with a green $+$ sign at its centroid location. 
}
\label{fig:stellar_render}
\end{figure}

Figure~\ref{fig:stellar_render} shows an image of galaxies for a small fraction of the simulated area (0.73\,arcmin$^{2}$) for LSST with four different stellar densities superimposed. The postage stamps for stars are simulated with the same convention as for galaxies (see \sect{imagesim} for details). This means that brighter stars will cover a larger fraction of the image; however, for optimization purposes, we limit the maximum radius of a given source to 30 arcsec.
We do not simulate saturation effects or other sensor or optical artifacts associated with very bright stars.
 
\begin{figure}[tbp]
\begin{center}

\includegraphics[width=0.75\textwidth]{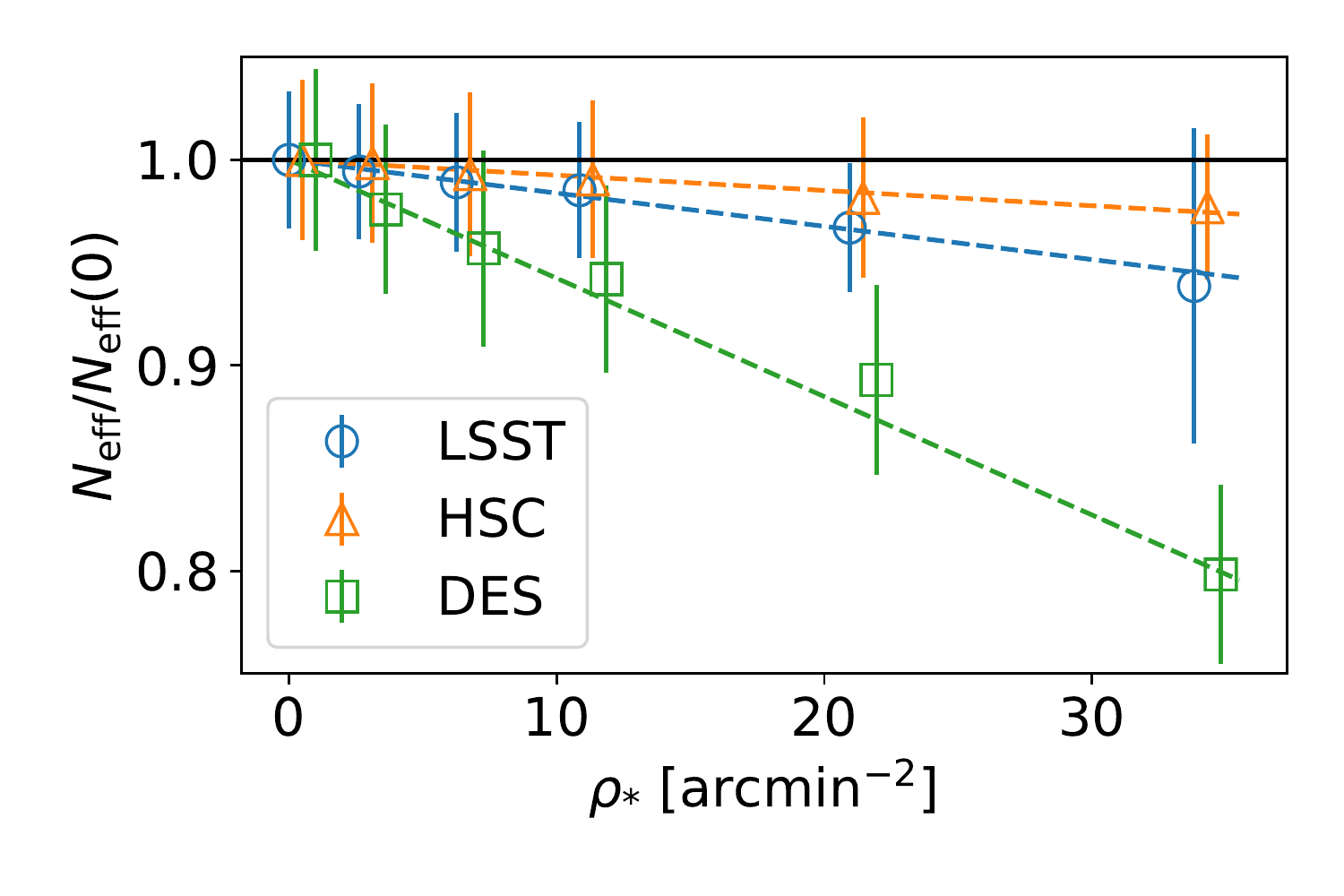}

\caption{Evolution of $\Neffgrpf(\rho_*)/\Neffgrpf(\rho_* = 0)$ with stellar density $\rho_{*}$ for LSST (blue circles), HSC (orange triangles), and DES (green squares) $i$ band. 
The dashed colored lines show the best fit for each case to the model in  \eqn{rho_neff}. The solid black line at 1 is added for reference. The different markers for a given value of $\rho_{*}$ are shifted with respect to each other for visualization purposes. We find a similar behavior for $r$ band.}
\label{fig:stellar_neff}
\end{center}
\end{figure}

The impact of the presence of stars on $\Neffgrpf$ is summarized in \fig{stellar_neff}. We  fit the values of $\Neffgrpf$ from this figure using the relationship
\begin{equation}
\Neffgrpf (\rho_*) = \Neffgrpf(0) \left(1+ A \rho_{*}\right) \;,
\label{eqn:rho_neff}
\end{equation}
where $\Neffgrpf(0)$ is the value of $\Neffgrpf$ with no stars, and $A$ has units of solid angle and is related to the average area lost due to the presence of a star. The general trend is that $\Neffgrpf/\Neffgrpf(0)$ decreases linearly with increasing stellar density. The best-fit values of $A$ are $(-5.8, -2.7 , -20.7)$ arcsec$^{2}$ for $i$-band LSST, HSC, and DES images, respectively. For $r$ band, we find $A=(-7.5, -3.1, -16.4)$ arcsec$^{2}$. 
The size of the PSF, rather than the depth of the survey, appears to play the most significant role in determining the fractional decrease in $\Neffgrpf$ with stellar density. 
The assumed PSF is smallest for HSC, for which we predict a smaller fractional decrease in $\Neffgrpf$ than for LSST. 
By far the largest predicted impact is for DES, which has the largest PSF of the three surveys.
We find that the impact of stars on $\Neffgrpf$ is similar for high- and low-purity galaxies, with purity defined with no stars present.

To determine whether statistical correlations between measures of overlapping stars and galaxies, or the higher shot noise due to the presence of the stars, is the more significant contributor to the decrease in $\Neffgrpf$, we compute  $\Neffgrpf$ in the sky-dominated limit ($\mu_{p} \sim b)$ for two values of stellar density. For $\rho_{*} \sim$ 2 arcmin$^{-2}$ and $\sim 10$ arcmin$^{-2}$, we find a relative decrease in $\Neffgrpf$ of 0.3\% and 0.4\%, respectively, due to only the increased shot noise of the overlapping stars (c.f. 0.2\% for galaxies only; see previous section). For comparison, the total reduction due to the presence of stars is 0.6\% and 1.5\% for $\rho_{*} \sim 2$ and 10  arcmin$^{-2}$. Therefore, the impact of source shot noise is higher than what we observed for galaxies only, and accounts for $\sim 25\%$ of the decrease in $\Neffgrpf$ due to stars. The remaining decrease comes from correlations in the measures of stars and galaxies. The lower fractional impact of correlations due to stars, compared to the case of galaxies only, is likely because stars have fewer free parameters (3) than galaxies (6). 

\begin{figure}[tbp]
\begin{center}

\includegraphics[width=0.9\textwidth]{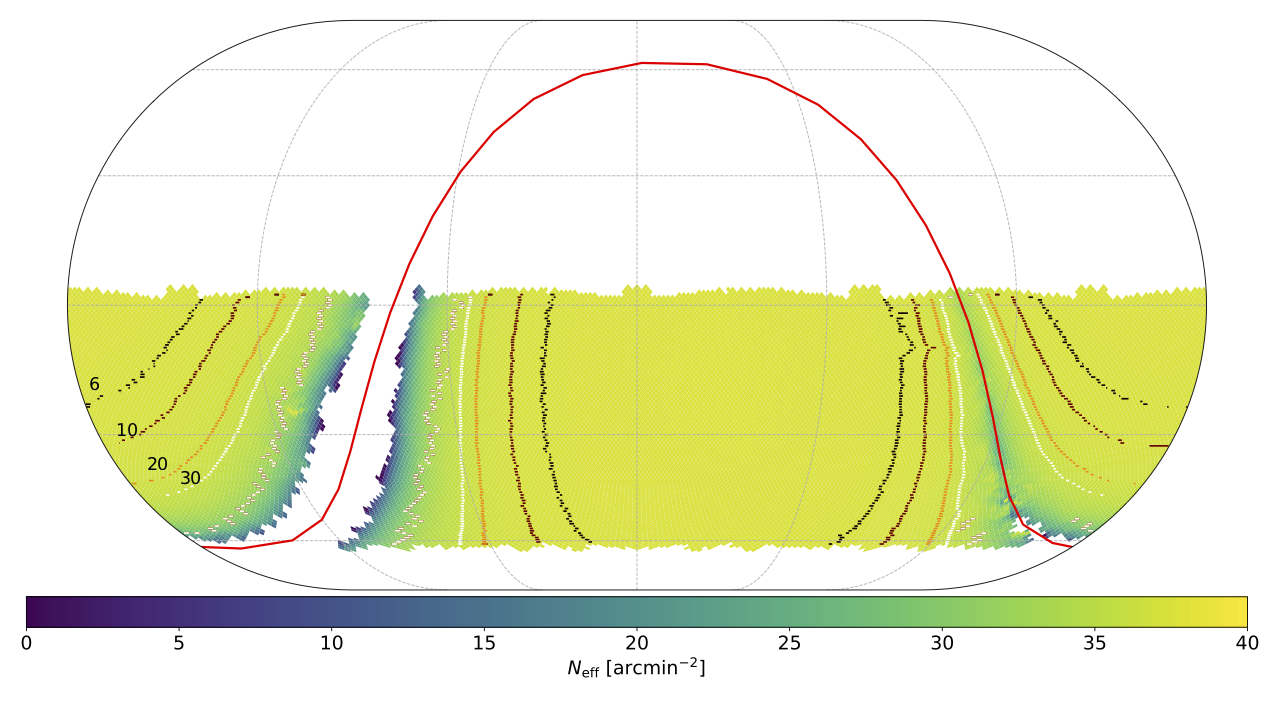}
\caption{Map of $\Neffgrpf(\rho_*)$ predicted from the fitted curves shown in  \fig{stellar_neff} for LSST $i$ band at full depth as a function of position on the sky in equatorial coordinates (RA, DEC). The contours correspond to four values of stellar density: $\rho_* = 6.0$, 10.0, 20.0, and 30.0 stars arcmin$^{-2}$, from furthest to closest to the galactic plane. 
This illustrates the area that would be lost if any of these thresholds were used as a maximum stellar density for analyzing the field. Similar behavior is found for $r$ band.}
\label{fig:neff_footprint}
\end{center}
\end{figure}

In addition to the loss of statistical sensitivity quantified with $\Neffgrpf$, we expect that regions of high stellar density will have an associated increase in systematic errors due to the more challenging star-galaxy separation and contamination of faint stars misclassified as galaxies.  In practice, an LSST weak-lensing analysis will likely be restricted to regions of low extinction~\cite{2018arXiv181200515L}, $E(B-V) < 0.2$, which roughly correspond to regions with low stellar density ($\rho_{*} < 10$ arcmin$^{-2}$).  
Using our fitted model for $\Neffgrpf(\rho_*)$, we predict $\Neffgrpf$ over the entire LSST survey area;  
the results are shown in \fig{neff_footprint} as a map of $\Neffgrpf$ for LSST $r$ band at full depth.    
We then quantify the effective total survey area available for a given maximum stellar density $\rho_{*,\text{max}}$ by dividing the sky into pixels of equal area $A_{\rm pix}$ indexed by $i$ and calculating
\begin{equation}
A_{\rm eff}(\rho_{*,\text{max}}) = A_{\rm pix}\, \sum_i\,
\frac{N_{{\rm eff}, i}}{\Neffgrpf(\rho_* = 0)} \; ,
\label{eqn:eff_area}
\end{equation}
where the sum is only over pixels $i$ with stellar density $\rho_{*,i} < \rho_{*, \text{max}}$ and $\Neffgrpf(\rho_* = 0)$ is the effective number of galaxies with no stars present. \eqn{eff_area} is a fair measure of the power of a survey only in the case of shape-noise-limited statistics; when sample variance is important, additional area becomes more valuable.
The results are summarized in \tab{effective_area}. For the areas used in ref.~\cite{DESC_SRD} (12,300 and 14,300 deg$^{2}$ for years 1 and 10, respectively, for LSST), the effective area would be over 99\% of the survey area but requires imaging regions with $\rho_{*, \text{max}} > 10$ arcmin$^{-2}$.

\begin{table}
\centering
\begin{tabular}{|cccc|}
\hline
$\rho_{*,\text{max}}$ [stars arcmin$^{-2}$] & $A_{\rm eff}$ [deg$^{2}$] &  $A_{\rm survey}$ [deg$^{2}$] & $A_{\rm eff}/A_{\rm survey}$\\
\hline
6 & 9055 & 9108 & 99.4\%\\ 
10 & 12082 & 12173 & 99.2\%\\
20 & 15006 & 15165 & 98.9\%\\
30 & 16342 & 16556 & 98.7\%\\
\hline
\end{tabular}
\caption{Effective and actual survey areas for LSST $i$ band, and ratio of the two (right column), for different values of the stellar density threshold $\rho_{*,\text{max}}$.}
\label{tab:effective_area}
\end{table}

\begin{figure}
    \centering
    \includegraphics[width=0.75\textwidth]{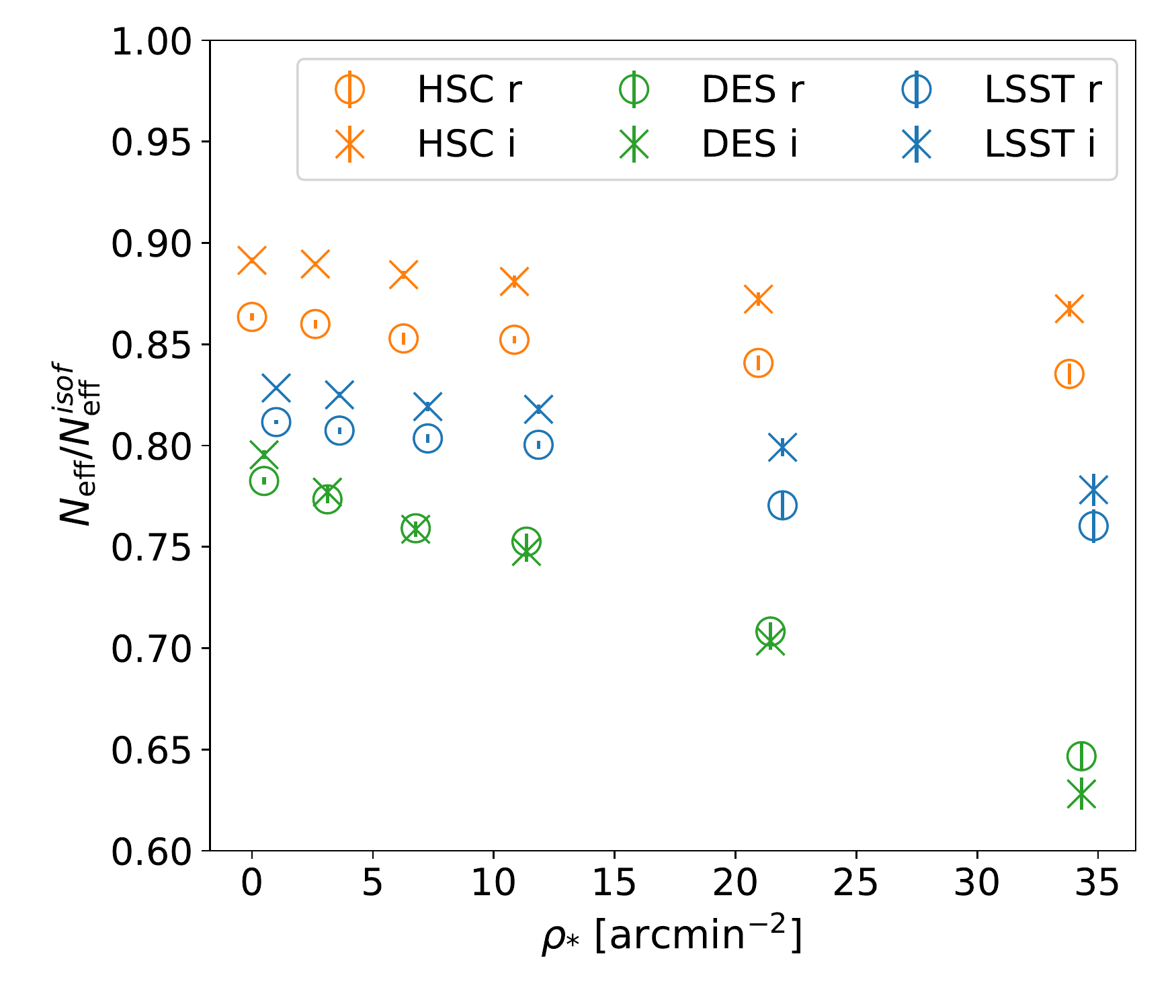}
    \caption{Loss in effective number density of galaxies due to blending as parameterized by the ratio $\Neffgrpf/\Neffisof$ for LSST (green), HSC (blue), and DES (orange) for both $r$ (open circles) and $i$ (crosses) bands, as a function of the stellar density, for the same fields used to generate the plots in~\fig{stellar_neff}.  $\Neffgrpf$ is calculated using grpf weights for galaxies with $\nu_{\rm grpf} >6$ -- i.e., including parameter correlations and source noise introduced by blending. $\Neffisof$ is calculated using isof weights for galaxies with $\nu_{\rm isof} > 6$ -- i.e., galaxies are treated as isolated. The markers for different surveys for a given value of $\rho_{*}$ are shifted with respect to each other for visualization purposes.}
    \label{fig:ratio_stellardensity}
\end{figure}

Another  way to quantify the impact of blending on the statistical sensitivity to cosmic shear is to compare the effective number density calculated with grpf weights for galaxies with $\nu_{\rm grpf} > 6$ ($\Neffgrpf$) and the effective number density when all galaxies are treated as isolated -- i.e., the effective number density calculated with isof weights for $\nu_{\rm isof} > 6$ ($\Neffisof$). The ratio $\Neffgrpf/\Neffisof$ indicates the fraction of signal lost due to the fact that detection and parameter estimation are both impacted when galaxy images overlap and parameters become correlated. The value of this ratio as a function of stellar density is shown in~\fig{ratio_stellardensity}. The dependence of $\Neffgrpf/\Neffisof$ on stellar density is similar to the dependence of $\Neffgrpf$ with stellar density, with the ratio  decreasing as stellar density increases due to increasing correlations as the number of objects in a group increases. Even in the case of only galaxies ($\rho_{*} =0$), the overlap between galaxies generates correlations and noise that translate into a $\sim 18\%$ loss for LSST ($\sim 13\%$ for HSC) in the total effective number density, compared to the case when they are isolated.  

The ratio $\Neffgrpf/\Neffisof$ is higher for HSC than for LSST, indicating that the impact of blends is less for  HSC than for LSST. This is due to the smaller PSF size for HSC compared to the predicted size for LSST. The ratio $\Neffgrpf/\Neffisof$ for DES is less than that for both LSST and HSC, driven mostly by the larger PSF size for DES.  Although the value of $\Neffisof$ is in practice unreachable because it requires every object to be detected in isolation, it provides  an upper limit on $\Neffgrpf$. 

These predictions are based on single-band images. Combining multiple bands, or external datasets from other surveys, can potentially break  degeneracies generated by blending, increasing the value of the ratio $\Neffgrpf/\Neffisof$ and mitigating the impact of blending.

\section{Results: Increase in pixel-noise bias due to blending}
\label{sec:Pixel-noise bias}

In this section, we first quantify pixel-noise bias for maximum-likelihood estimators of galaxy shape parameters for overlapping galaxy images, as a function of the distance between the galaxy centroids and the relative orientation of the galaxies, for fixed sizes, ellipticities, and fluxes of the two galaxies (\sect{noisebias_distance}). We then use the simulated full-depth LSST observations to quantify the expected increase in noise bias on galaxy shape measurements due to overlapping sources and the impact on cosmic shear (\sect{noisebias_shear} through \sect{noise_bias_on_shear}). Finally, in \sect{bias_redshift}, we examine the dependence of multiplicative shear bias on redshift.

\subsection{Dependence of noise bias on galaxy separation and relative orientation}
\label{sec:noisebias_distance}

\begin{figure}[tbp]
\begin{center}
\includegraphics[width=0.9\textwidth]{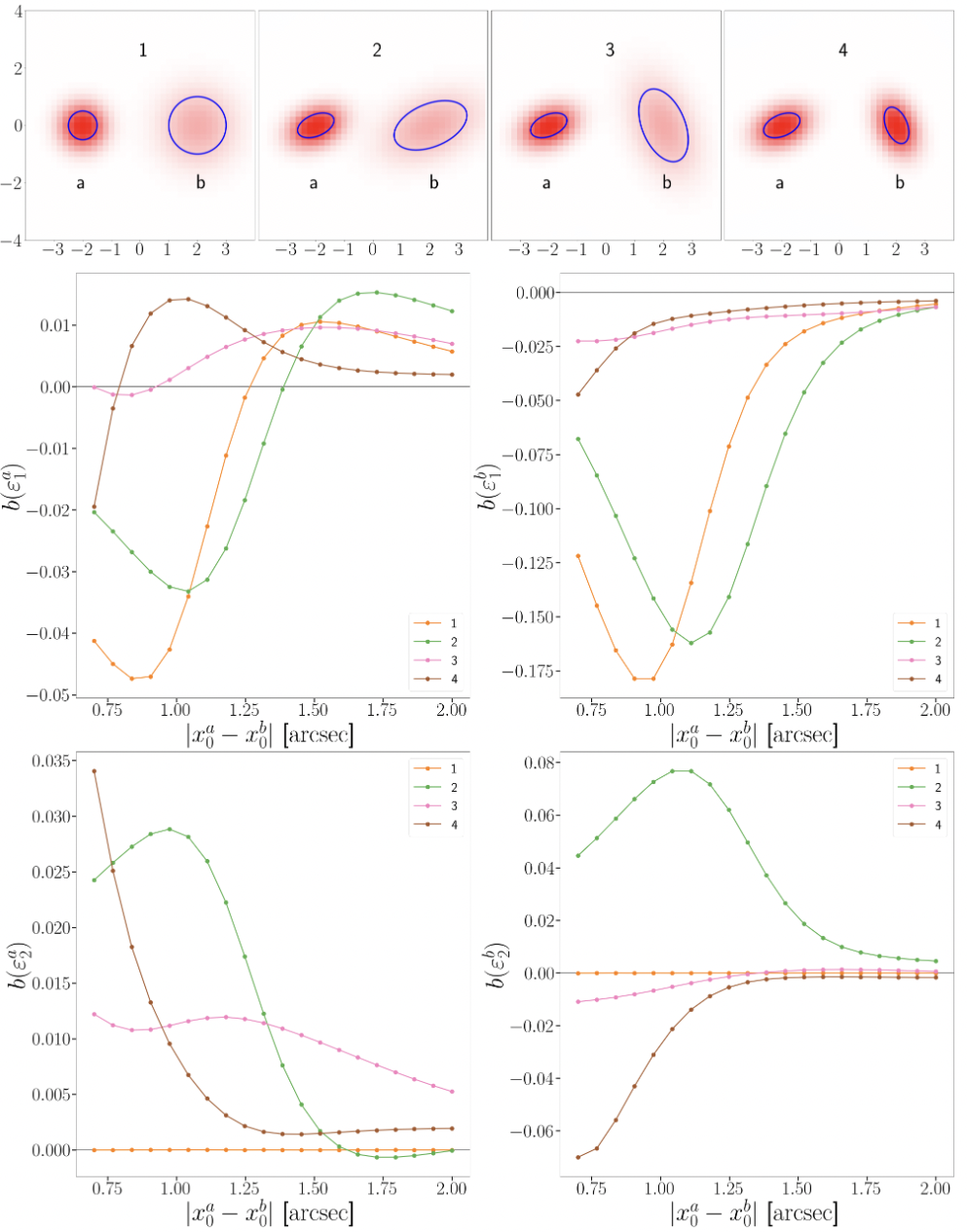}
\caption{Fisher predictions for dependence of predicted noise bias for ellipticity component $\varepsilon_i^j$, 
as a function of the distance between the centroids of the two galaxies, $|x_0^\textsf{b} - x_0^\textsf{a}|$.  
The subscript $i$ on $\varepsilon_i^j$ denotes ellipticity component 1 or 2, while the superscript $j$ denotes galaxy $\textsf{a}$ or $\textsf{b}$ with parameters defined in \tab{galaxy_pair_parameters} and illustrated in the top row of images for a separation of 4 arcsec. The scale for the vertical and horizontal axes for the images in the top row is arcseconds.
}
\label{fig:two_gal_vary_sep}
\end{center}
\end{figure}

We begin to quantify the impact of overlapping galaxy images on noise bias by considering the bias in shape parameters for two overlapping galaxies with centroid positions $x_0^\textsf{a} = -x_0^\textsf{b}$ and $y_0^\textsf{a} = y_0^\textsf{b} = 0$,
where the superscripts $\textsf{a}$ and $\textsf{b}$ denote the
left and right galaxy in each of the four images 
shown at the top of \fig{two_gal_vary_sep}.
We use a Gaussian PSF with HLR=$0.35\,$ arcsec, for 0.2-arcsec pixels in a 40 pixel by 40 pixel image. The fluxes of the two galaxies in a pair are the same. The HLR, ellipticity, and $\nu_{\rm sky}$ for each of the two galaxies are given in \tab{galaxy_pair_parameters}. 
The $\nu_{\rm sky}$ value shown for galaxy $\textsf{a}$ is used to calculate the corresponding pixel noise value when galaxy $\textsf{a}$ is the only source in the image; this value of the pixel noise is then used to calculate the $\nu_{\rm sky}$ value for galaxy $\textsf{b}$ when it is the only source in the image.  
In the lower four plots in \fig{two_gal_vary_sep}, we plot the predicted noise bias for ellipticity component $\varepsilon_i^j$, $j=\textsf{a},\,\textsf{b}$,
as a function of the distance between the centroids of the two galaxies for  $|x_0^\textsf{b} - x_0^\textsf{a}| > 0.7$\,arcsec.\footnote{The predicted noise biases for pair 4, in which the two galaxy images differ only in their ellipticity orientation, become very large for $\varepsilon^j_2$ for centroid separations \textit{less} than $0.7$\,arcsec. 
Hence, we restrict the range of separation to better illustrate the variation.}
The color of each curve corresponds to one of the four pairs. 

We see that the magnitude and sign of the noise bias on each ellipticity component depend in a complex way on the ellipticity, relative orientation, and separation of the two galaxies. 
The dependence is even more complex if the relative flux and $\nu_{\rm sky}$ of the overlapping galaxies is varied as well. 
To estimate the expected noise bias on the shear estimator $\langle \epsv'\rangle$ (\eqn{estimator}), we use a simulation of a portion of the LSST survey, described in the next section.

\begin{table}[tbp]
\centering
\begin{tabular}{|ccccrc|}
\hline
Galaxy Pair & Galaxy & HLR [arcsec] & $\varepsilon_1$ & $\varepsilon_2$ & $\nu_{\rm sky}$\\
\hline
1 & \textsf{a} & 0.5 & 0.0 & 0.0 & 20 \\
     & \textsf{b} & 1.0 & 0.0 & 0.0 & 12 \\
\hline
2  & \textsf{a} & 0.5 & $+0.2$ &  $+0.2$ & 20 \\
     & \textsf{b} & 1.0 & $+0.2$ &  $+0.2$ & 12  \\
\hline 
3  & \textsf{a} & 0.5 & $+0.2$ & $+0.2$ & 20\\
   & \textsf{b} & 1.0 & $-0.2$ & $-0.2$ & 12 \\
\hline
4  & \textsf{a} & 0.5 & $+0.2$ & $+0.2$ & 20 \\
     & \textsf{b} & 0.5 & $-0.2$ & $-0.2$ & 20 \\
\hline
\end{tabular}
\caption{True parameter values for the four simulated pairs of galaxies, illustrated at the top of \fig{two_gal_vary_sep}.
}
\label{tab:galaxy_pair_parameters}
\end{table}

\subsection{Quantifying noise bias on shear estimator}
\label{sec:noisebias_shear}
 
As discussed in sec.\,\ref{sec:shapeshear}, 
if intrinsic galaxy ellipticities $\epsv$ are uniformly distributed
in angle, the expectation value of the sheared galaxy shape
$\epsv'$, for fixed true shear $\gv$, is an estimator of shear:  
$\langle \epsv'\rangle = \gv$ (\eqn{estimator}).
By definition, the MLE for the sheared galaxy shape, averaged over
noise realizations for fixed PSF and galaxy parameters and fixed
SNR, is equal to the true value plus the noise bias $b_g$:
$$\hat\epsv' = \epsv' + b_g\,.$$
Taking the expectation value of each side over an ensemble of galaxies with varying parameters and $\langle \epsv' \rangle = \gv$, we have 
$$\langle \hat\epsv'\rangle = \langle\epsv'\rangle + \langle b_g \rangle = \gv + \langle b_g \rangle\;.$$
If we parameterize the expectation value of the shear estimator  in terms of multiplicative and additive biases $m$ and $c$,  
$$\langle \hat\epsv'\rangle = \gv (1 + m) + c\;,$$ 
then 
$\gv +  \langle b_g \rangle = \gv (1 + m) + c$ or  
\begin{equation}
\langle b_g \rangle = m \gv + c.
\label{eqn:shear_bias}
\end{equation} 

We use the simulated full-depth LSST observations described in sec.\,\ref{sec:imagesim} with {sheared galaxy shapes} and the Fisher predictions for $b_g$ to measure the shear bias parameters $m$ and $c$  according to \eqn{shear_bias}, for samples with blending on and with blending off, thereby quantifying the expected increase in noise bias on shear due to overlapping sources.

\subsection{Definition of ``lensing sample'' and ``accurately detected and deblended" galaxies}
\label{sec:sample_definitions}

Although we simulate images to full depth, we must identify sources that are likely to be detected, and for which the Fisher forecasts for noise bias are expected to be valid and relevant for ML fits -- in other words, sources that have sufficiently high SNR and are not too small compared to the PSF, and are sufficiently separated from neighboring sources so that they could be recognized as a separate object. 

Rather than using the traditionally defined LSST ``gold'' sample of galaxies (sources with $i < 25.3$), we define a set of galaxies, which we call the {``lensing sample''}, that satisfy the following criteria on SNR and galaxy size: 
$\nu_{\rm grpf} > 6$ and $\sigma_- > 0.2$\,arcsec. 
The requirement on $\nu_{\rm grpf}$ is the same as that used to define ``detectable'' galaxies in \sect{snr_purity}.
As described in \app{dep_on_size_ell}, the Fisher predictions for noise bias increase rapidly for galaxies that are small compared to the PSF.

\begin{figure}[tbp]
\begin{center}
\includegraphics[width=0.6\textwidth]{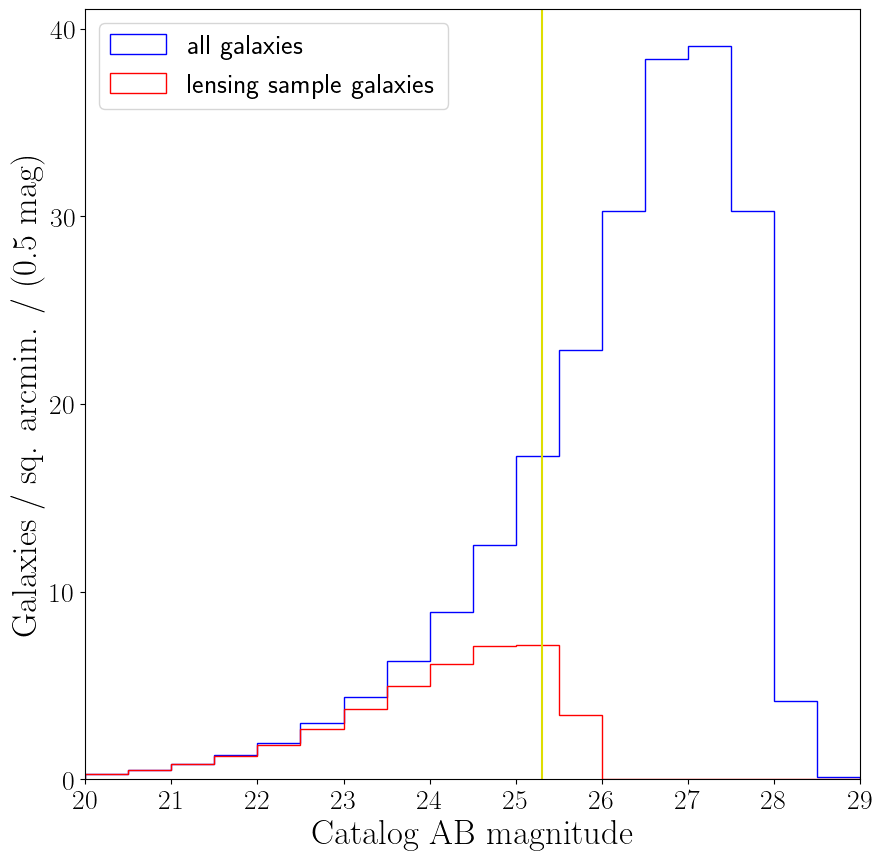}
\caption{Histogram of $i$-band magnitude for all galaxies in full-depth LSST simulation with 
(blue histogram), for galaxies with $\nu_{\rm grpf} > 6$ and $\sigma_- > 0.2$\,arcsec (lensing sample, red histogram), and for galaxies with $i < 25.3$ (gold sample, to the left of gold vertical line).}
\label{fig:i_good_gold}
\end{center}
\end{figure}

In \fig{i_good_gold}, we show the distribution of $i$-band magnitude for all simulated galaxies and for the 18.1\% of these galaxies that satisfy the criteria for {the lensing sample}.
The LSST ``gold'' sample corresponds to the 22.5\% of galaxies in the blue histogram that lie to the left of the vertical gold line in the figure. 

We next use the output of Source Extractor (SE) to define a set of simulated galaxies that are detected, and then divide these galaxies into those for which overlaps with other galaxies are accurately characterized -- or not -- as described in more detail below. 
We use the ``default'' SE settings given in \app{sextractor_settings}, which are optimized for extended sources; the image is convolved with a kernel smaller than the PSF, and the SNR threshold is set relatively low but more than one contiguous pixel above threshold is required. 
After running SE on the image, we match each detected object to the true object whose centroid is closest to that of the detected object; 
the closest true object for each detected object is called a ``primary match'' and the set of primary matches is labeled as ``detected''. 
A set of objects is labeled as ``close overlaps'' if the true centroid of at least one \textit{non}-primary (true) object lies within one unit of the detected centroid of a detected object, where the unit of distance is the sum of the PSF-convolved size of the non-primary true object and the size of the detected object as measured by SE. 
We use this definition of close overlaps as a proxy for identifying overlapping objects that a detection algorithm would not, in general, be able to distinguish as separate objects. 

Other galaxies of interest are those that are members of a blended group with a non-invertible (equilibrated) Fisher matrix. These are the groups for which at least one galaxy was dropped from the Fisher analysis, as described in  \sect{statsensitivity}.  Given the non-robustness of the matrix inversion for these groups, we drop this very small fraction of galaxies.

\subsection{Sample used for noise bias estimates}
\label{sec:bias_sample}

The total number of simulated galaxies in our catalogue is $\sim 800\mathrm{k}$. Of the $18.1\%$ of simulated galaxies that satisfy the criteria for the \textit{lensing} sample ($\nu_{\rm grpf} > 6$ and $\sigma_- > 0.2$\,arcsec), 93.2\% are detected -- 77.9\% with no close overlaps and 15.4\% with close overlaps; 6.8\% are not detected -- 4.8\% with no close overlaps and 2.0\% with close overlaps. 

In order to approximate the sample of galaxies that could be accurately detected and deblended, and apply the Fisher formalism to predict noise bias for this sample, we select the galaxies in the lensing sample that are detected, have no close overlaps, and do not belong to a group with dropped sources. This results in a sample of $\sim$113k galaxies, which is $\sim$14\% of the total number of galaxies.
Since an accurate hypothesis for a ML fit is not possible for galaxies that cannot be accurately detected and/or deblended, the Fisher formalism is not relevant for those galaxies. 
Galaxies that cannot be accurately detected and/or deblended can produce other types of shear bias \cite{2017arXiv170202600H,2016ApJ...816...11D}; however, these issues are not addressed in this study. Of these 113k selected galaxies, $85.6\%$ are part of a group with more than one galaxy.

\subsection{Weighted mean of shear biases}
\label{sec:central_tendency}
We find that the predicted impact of noise bias on shear estimation depends sensitively on whether we weight the Fisher estimates of noise bias using their estimated uncertainties, and whether we clip the sample to remove extreme outliers. Ultimately, we decided to use an (unclipped) \textit{weighted mean} of the noise bias for ensemble averaging since it is the most useful for real analyses of shear (e.g., calculating two-point shear correlation functions).
We calculate the following weight for each noise bias estimate: 
\begin{equation}
    w_{i,j} = \frac{1}{\sigma_{s,j}^{2} + \sigma^{2}_{m,i,j}} , 
\label{eqn:weighting-scheme}
\end{equation}
where $i$ is the galaxy index, $\sigma^{2}_{s,j}$ is the shape noise of the $j$th component of shear (i.e., $\sigma_{s,j}^2=\langle \varepsilon^{2}_{j} \rangle$),  
and $\sigma^2_{m,i,j}$ is the Fisher-predicted uncertainty on the $j$th shear component for the $i$th galaxy. Based on these weights, we define the \textit{weighted mean} of our noise bias sample as 
\begin{equation}
    \hat{b}_{w,j} = \frac{\sum_{i=1}^{n} w_{i,j} x_{i,j}}{\sum_{i=1}^{n} w_{i,j}}.
\label{eqn:weighted-mean}
\end{equation}

\subsection{Estimation of multiplicative and additive noise bias on shear \label{sec:noise_bias_on_shear}}
\begin{figure}[tbp]
   \centering
\begin{subfigure}[b]{0.475\textwidth}   
    \centering 
    \includegraphics[width=\textwidth]{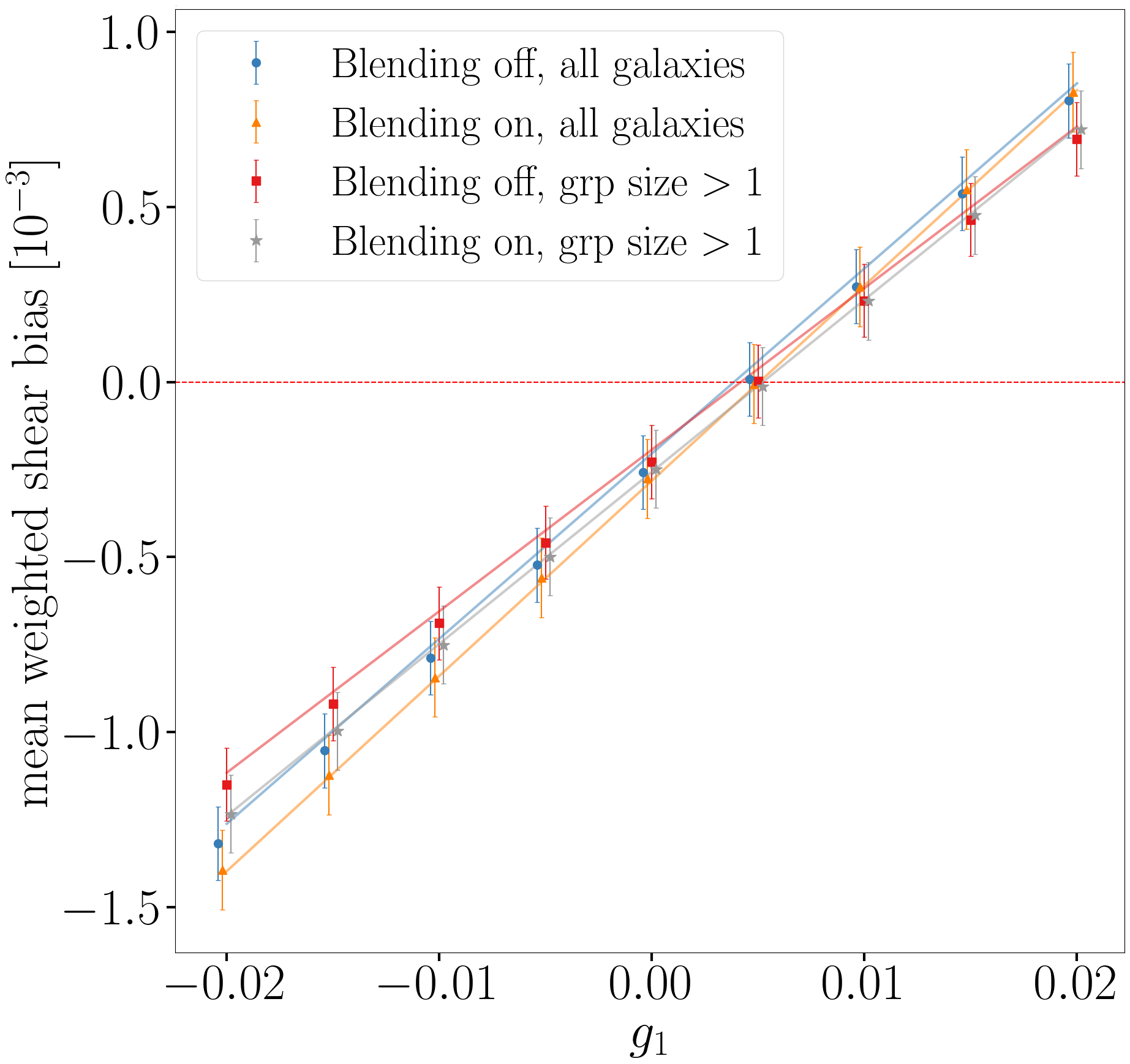}
    \end{subfigure}
    \quad
\begin{subfigure}[b]{0.463\textwidth}   
    \centering 
    \includegraphics[width=\textwidth]{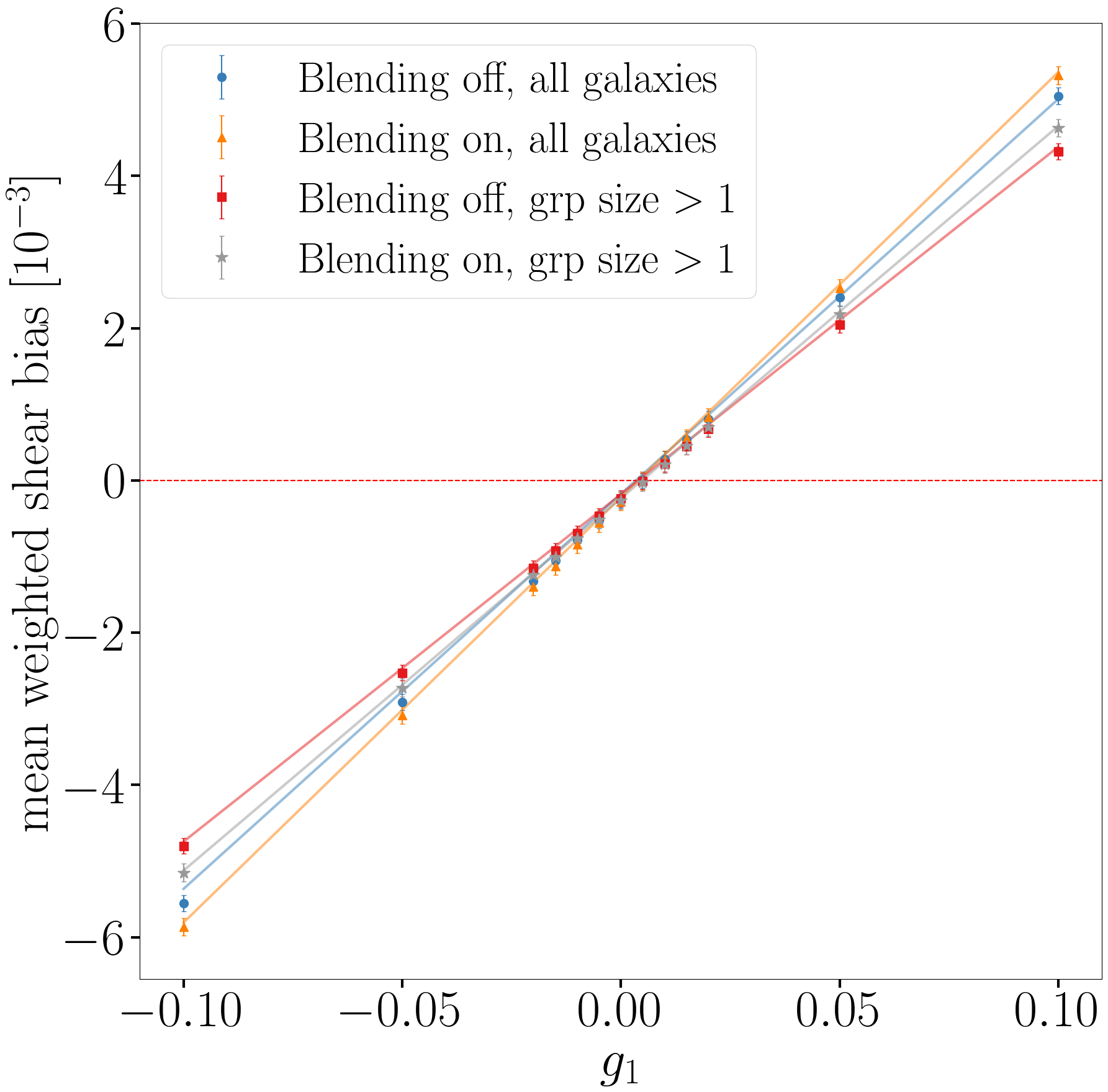}
\end{subfigure}
\caption{Weighted mean noise bias for $g_1$ ranging from $-0.020$ to $+0.020$ in steps of $0.005$ (left) and including the four ``large'' applied shears (right) and $g_{2} = 0 $, with blending off or on, and for all galaxies in the sample or only galaxies that are members of groups of size $>1$.
Each data point corresponds to the ensemble average of the predicted noise bias for a sample of galaxies in the lensing sample that are expected to be accurately detected and deblended. 
The error bars correspond to the uncertainties estimated with a bootstrap technique.
The lines in the figures correspond to a linear fit in the correlated data points; the results are shown in \tab{bias_results}. In the left figure, we introduce a small horizontal offset on each datapoint for visual clarity. The points with the red square marker have no offset. }
\label{fig:noise_bias_good_det}
\end{figure}

To measure multiplicative noise bias on shear, we use the \textsc{GalSim} package to apply a constant shear to all galaxies in the \textsc{CatSim} catalog.
We produce thirteen samples, denoted $S_{1}, S_{2}, ..., S_{13}$, each with a different value of applied shear $g_1$ but with $g_{2} = 0$ for all samples. The samples include four large values of applied shear $g_{1}=\pm 0.1, \pm 0.05$, as well as $g_{1}$ ranging from $-0.020$ to $+0.020$ in steps of $0.005$, and correspond to the data points in  \fig{noise_bias_good_det}. 
The applied shear can subtly affect some galaxy parameters on which the selection of accurately detected and deblended objects in the lensing sample depends.\footnote{Approximately 10\% of the galaxies that pass the selection criteria for \textit{at least one} of the applied shears, do not pass the criteria for \textit{all} of the applied shears. } 
Therefore, to avoid ``selection bias'' in our analysis, for each applied shear, we use the same set of galaxies -- specifically, the $\sim$ 113k galaxies that are classified as accurately detected and deblended in the sample with zero applied shear.

For each galaxy sample $S_{i}$, we compute the weighted shear bias using \eqn{weighted-mean} for the $j=1$ shear component. We plot these values in \fig{noise_bias_good_det} as a function of the applied shear component $g_1$, for blending off and blending on, for all galaxies and for only those that are members of a group with more than one galaxy, both for the accurately detected and deblended sample.

\begin{table}[tbp]
\centering
\begin{tabular}{|c|c|cc|}
\hline
Sample & Blending & Multiplicative bias $m$ & Additive bias $c$\\
\hline
\vspace{-0.3cm} & & & \\
All & off & $(5.29 \pm 0.03) \times 10^{-2}$ & $(-2.0 \pm 0.9) \times 10^{-4}$\\
\vspace{-0.3cm} & & & \\
All & on & $(5.58 \pm 0.03) \times 10^{-2}$ & $(-2.8 \pm 1.1) \times 10^{-4}$\\
\vspace{-0.3cm} & & & \\
Group & off & $(4.61 \pm 0.03) \times 10^{-2}$ & $(-1.9 \pm 0.9) \times 10^{-4}$\\
\vspace{-0.3cm} & & & \\
Group & on & $(4.91 \pm 0.03) \times 10^{-2}$ & $(-2.6 \pm 1.1 ) \times 10^{-4}$\\
\hline
\end{tabular}
\caption{ Multiplicative ($m$) and additive ($c$) noise bias on shear for blending off and blending on, for all galaxies in the sample (All) and for the subset of galaxies that belong to a group with more than one galaxy (Group) with applied shear $g_{1}$ ranging from $-0.02$ to $0.02$ and $g_{2}=0$. The values correspond to the slope and intercept of a linear fit to the correlated data points shown in the left figure of \fig{noise_bias_good_det}, which themselves correspond to the weighted mean predicted noise bias for a sample of galaxies in the lensing sample that are expected to be accurately detected and deblended.
}
\label{tab:bias_results}
\end{table}

We extract $m$ and $c$ from a linear fit to the values in \fig{noise_bias_good_det}, taking into account correlations between the thirteen points using a bootstrap method. The results are shown in \fig{noise_bias_good_det} and \tab{bias_results} for the samples with blending off and blending on. 

We see that the magnitudes of the biases  and the multiplicative shear bias are larger for blending on than for blending off as shown in table \ref{tab:bias_results}. The observed additive bias is due to an asymmetry in the $\sim10\%$ of values furthest from the core of the distribution of weighted biases. Also, the multiplicative noise biases from the subset of galaxies that belong to groups with more than one galaxy are \textit{lower} than the corresponding biases for the whole sample. This can be attributed to the significant number of small and noisy isolated galaxies that, on average, have a large predicted noise-bias value.\footnote{We also repeated this procedure applying a symmetric shear $g_{1} = g_{2}$. We find no significant difference in the values of the multiplicative and additive biases obtained in these symmetric case and those reported in table \ref{tab:bias_results}.}

\subsection{Dependence of multiplicative bias on redshift} \label{sec:bias_redshift}

As documented in secs.\,5.2, D2.1 and D2.3 in version 1 of the 
LSST DESC Science Requirements Document (SRD) \cite{DESC_SRD},
the requirement we must meet for the 10-year DESC weak lensing ($3\times2$-point) analysis is that the \textit{total systematic uncertainty} in the redshift-dependent shear calibration not exceed $0.003$.
The SRD analysis uses five photometric redshift bins, each 0.2 units wide, in the redshift range $0.2 \leq z \leq 1.2$.
A linear parameterization is suggested for the redshift dependence of the multiplicative shear bias, $m(z)$: 
\begin{equation}
    m\left(z\right) = m_{0} \left(\frac{2z - z_{\max}}{z_{\max}} \right) + m_{\rm avg}\;,
\label{eqn:tomo-z-equation}
\end{equation}
where $z_{\max}$ is the redshift value at the center of the highest redshift bin, 
$m_{\rm avg} = m(z_{\max}/2)$ is the average value of $m$ in the range $z \in \left[0,z_{\max}\right]$, and 
$2m_{0} = m(z_{\max}) - m(0)$ is the total variation in $m$ in the range $z \in \left[0,z_{\max}\right]$.
The 0.003 requirement is then on the total uncertainty on $m_0$ due to all contributions to multiplicative shear bias. In general, there might be higher-order dependence on the redshift $z$ for the multiplicative shear bias $m(z)$. Since there is no explicit requirement on $m(z)$ in the case that it is non-linear, we translate the requirement of $2\left(0.003\right) = 0.006$ on the uncertainty on $2m_{0} = m(z_{\max}) - m(0)$ onto a requirement on the uncertainty on the maximum span of $m(z)$ over the range $z \in \left[0,1.2\right]$.

\begin{figure}[ht]
\centering
\includegraphics[width=0.6\textwidth]{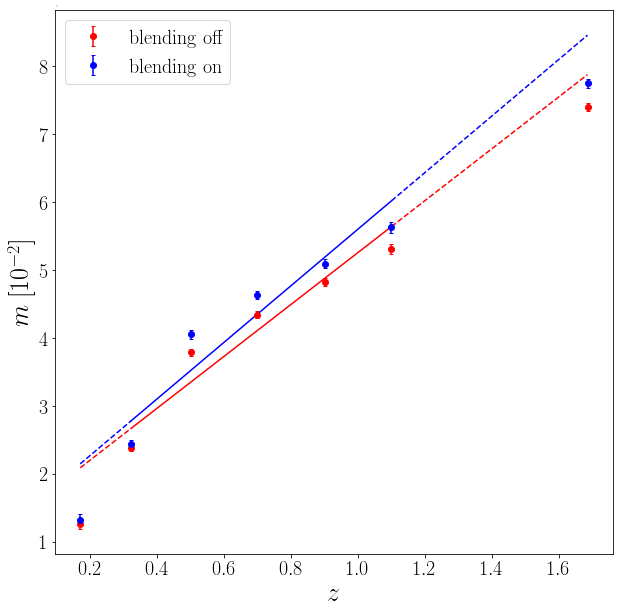}
\caption{Predicted multiplicative shear bias $m$ due to pixel noise as a function of redshift $z$, for blending off (red) and blending on (blue) using the weighted means.
The points with error bars represent the value of $m$ determined for the samples of galaxies in five redshift bins of width 0.2, between  0.0 and 1.2;  the right-most red and blue points correspond to galaxies with $z>1.2$. 
The horizontal position of each point corresponds to the median value of $z$ for that sample. 
The lines correspond to fits to \eqn{tomo-z-equation}.
The points corresponding to the $\left[0.0,0.2\right]$ and overflow bins in $z$ are not included in the fits.
}
\label{fig:tomo-z-result}
\end{figure}

To estimate $m_{\rm avg}$ and $m_0$, we measure $m$ using the same bootstrap and fitting technique described in the previous section, for seven non-overlapping subsets of galaxies from the accurately detected and deblended sample: 
galaxies with redshifts lying in six bins spaced by 0.2 in the range $ 0.0 \leq z \leq 1.2.$, and one ``overflow'' bin ($z>1.2$). 
The values of $m$ and their uncertainties are shown in \fig{tomo-z-result} for blending off (red points) and blending on (blue points), and using the weighted mean for ensemble averages, as a function of the median redshift for each bin. These results agree with our expectation that the multiplicative bias increases with redshift due to decreasing average galaxy size and SNR values.

The red and blue lines show the results of a least-squares fit\footnote{We find that the correlation between multiplicative biases in different redshift bins are negligible ($< 0.1$) and thus ignore them in the fits.} to \eqn{tomo-z-equation} for each set of points. The values of $m_{\rm avg}$ and $m_0$ for this case are displayed in the first two rows of table \ref{tab:tomo_bias_results}. The multiplicative shear bias $m(z)$ derived from the weighted mean is non-linear in $z$; see \fig{tomo-z-result}. The corresponding values in table \ref{tab:tomo_bias_results} indicate that $m_{0}$ is an order of magnitude higher than the SRD requirement.  

\begin{table}[tbp]
\centering
\begin{tabular}{|c|ccc|}
\hline
\vspace{-0.3cm}& & & \\
Sample & $m_{0}$ & $m_{\rm avg}$ & Correlation \\
\hline
\vspace{-0.3cm} & & &  \\
Blending off & $(2.1 \pm 0.4)\times 10^{-2}$ & $(3.5 \pm 0.6) \times 10^{-2}$ & $-0.5$ \\ 
\vspace{-0.3cm} & & & \\
Blending on & $(2.3 \pm 0.4) \times 10^{-2}$ & $(3.7 \pm 0.7) \times 10^{-2}$ & $-0.5$ \\ 
\hline
\end{tabular}
\caption{ Values of $m_{0}$ and $m_{\rm avg}$ obtained after fitting \eqn{tomo-z-equation} to the multiplicative shear biases across different redshift bins in \fig{tomo-z-result} for blending off or on and using the weighted mean. } 
\label{tab:tomo_bias_results}
\end{table}

\section{Conclusions}
\label{sec:conclusions}

We have presented a framework for analyzing two impacts of blending on measurements of cosmic shear in single-band images:
the decrease in the effective number density of galaxies ($\Neffgrpf$) due to the increase in statistical uncertainty on shape parameters, and
the increase in pixel-noise bias for shape parameters.   
Although the Fisher formalism that is used makes no explicit assumption about detection or deblending algorithms, in effect it assumes that both the true number of galaxies and the model for the galaxy profiles are known. 
Therefore, the predicted values of $\Neffgrpf$ are the maximum possible values for the selected sample. 

We introduce a measure of signal-to-noise ratio ($\nu_{\rm grpf}$) that accounts for the increase in statistical uncertainty on parameter estimates due to overlapping images.
We then determine $\nu_{\rm grpf}$ thresholds to use as a surrogate for detection by measuring the detection efficiency for SourceExtractor applied to noisy simulated images, as a function of $\nu_{\rm grpf}$, for three different full-depth surveys: DES, HSC, and LSST. 

\subsection{Impacts of blending on $\Neffgrpf$}
The predicted values of $\Neffgrpf$ are shown in~\tab{summary} 
for DES, HSC, and LSST, in the $i$  and $r$ bands, for galaxies with $\nu_{\rm grpf}$ above the detection threshold. 
We predict a value of  $39.4$ galaxies per arcmin$^2$ for $\Neffgrpf$ in the LSST $r$ band.
One way to quantify the impact of blending is through the ratio $\Neffpur/\Neffgrpf$, where $\Neffpur$ is the contribution to $\Neffgrpf$ from only high-purity galaxies ($\rho \ge 0.98$). 
We find that this ratio is $\sim 0.74$ for HSC and DES $i$ and $r$ bands, and $\sim 0.62$ for LSST $i$ and $r$ bands, quantifying the expectation that blending has a greater impact at LSST depths. The impact is also quantified with the ratio $\Neffgrpf/\Neffisof$, which represents the ratio of effective number density of galaxies taking into account statistical correlations due to blending, compared to the effective number density if galaxies were isolated. Since this ratio is $\sim 82\%$ (for galaxies only) for LSST,  $\sim 18\%$ of the reduction in statistical sensitivity is due to correlations between galaxies in single-band images. This effect can be mitigated by combining information from multiple bands and/or using external datasets. We find that the effective number density decreases due to source shot noise by $\sim 0.2\%$ (for galaxies only), validating the use of the sky-dominated limit, where source shot noise is neglected, for ground-based surveys.

To check the robustness of our technique for estimating $\Neffgrpf$, we emulate the criteria applied by the DES collaboration for Science Verification data~\cite{2016MNRAS.460.2245J} and by the HSC collaboration for Subaru HSC-SSP survey year 1 data~\cite{2018PASJ...70S..25M} and find that our predictions for $\Neffgrpf$ for the selected simulated objects are consistent with the values measured by DES and HSC (see \sect{prior_studies}).

We present the impact of a range of stellar densities $\rho_*$ on the statistical sensitivity of shear measurements. 
We find that $\Neffgrpf$ decreases linearly with increasing stellar density as described in \eqn{rho_neff}, with a slope value that is correlated with PSF size; see \fig{stellar_neff}.
We find that the impact of stars on $\Neffgrpf$ is similar for high- and low-purity galaxies in the range of stellar densities we explored (from $\rho_{*}=0$ to $\rho_{*}~\sim 30$ arcmin$^{-2}$). We quantify the loss in the effective number density due to the presence of overlapping sources by plotting the ratio $\Neffgrpf/\Neffisof$ as a function of stellar density; see~\fig{ratio_stellardensity}. The presence of stars further lowers the effective number density, leading to a decrease in the ratio $\Neffgrpf/\Neffisof$ of up to $1.5\%$ for $\rho_{*} \sim 10$ arcmin$^{-2}$. 
The impact of source shot noise is higher than what we observed for galaxies only, and accounts for $\sim 25\%$ of the decrease in $\Neffgrpf$ due to stars. The remaining decrease comes from correlations in the measures of stars and galaxies. The lower fractional impact of correlations due to stars, compared to the case of galaxies only, is likely because stars have fewer free parameters (3) than galaxies (6).

We use the model for $\Neffgrpf(\rho_*)$ in \eqn{rho_neff} to predict the distribution of $\Neffgrpf$ on the sky given the stellar density distribution expected for the LSST, as described in the \textsc{CatSim} catalog. 
As expected, the regions close to the galactic plane do not contribute significantly to the integrated $\Neffgrpf$; see \fig{neff_footprint}.  
We calculate the actual survey area and the effective survey area -- i.e., the equivalent area for $\Neffgrpf(\rho_* = 0)$ --  for several stellar density thresholds; see \tab{effective_area}.

These results do not include the impact of  misclassifying stars as galaxies, which can introduce a systematic effect in shape estimation. 

\subsection{Impacts of blending on shear bias due to pixel noise}
We use the Fisher formalism to study the impact of blending on shape measurement and cosmic shear bias due to pixel noise for maximum-likelihood estimators. 
We compare pixel-noise bias for two different commonly used shear estimators ($\chiv$ and $\epsv$), for isolated and blended objects.
We measure the resulting multiplicative shear bias as a function of redshift for different measures of “ensemble average”. 
We show that the sign and magnitude of the predicted pixel-noise bias depends on the shear estimator that is used; see \fig{size_bias_eps2=0}. 
We define a set of criteria to select a sample of galaxies that would be typically selected for a lensing analysis, and that are likely to be accurately detected.   
We find that the ``mean'' noise bias for the sample of selected galaxies (and the multiplicative shear calibration bias calculated from the ensemble averages) has a significant dependence on how we calculate the ensemble average. 
We calculate the multiplicative shear bias for six redshift ranges between 0 and 1.2 and find that the bias increases with redshift due to decreasing average size and SNR. 
 The redshift-dependent multiplicative shear bias derived from the weighted sample mean biases is an order of magnitude greater than the LSST dark-energy requirements; see \tab{tomo_bias_results}. 
 
 Based on the magnitude of the estimated biases and the dependence on many factors including redshift-dependent galaxy properties, we conclude that the noise bias for ML shape estimators cannot be robustly estimated based only on simulations at the sensitivity required for LSST measurements of cosmic shear. 

\subsection{Limitations of this analysis and future prospects} 
In this study, we focused on only two impacts of blending on cosmic shear measurements: the loss of statistical power and the increase in pixel-noise bias. 
We did not analyze other systematic effects in shear measurements due to blending that depend on the algorithms used for detection, flux assignment, and measurement.  
For example, the impact of blending on  measurements of galaxy shapes and photometric redshifts will depend on the particular  algorithms used.
Although undetectable sources are included in the simulations, we did not directly address systematic biases in  measured shapes and fluxes of detected objects due to overlapping sources that are \emph{not} detected. 
These systematic effects will lead to further decreases in the effective number density compared to those predicted in this paper. 

Since pixel-noise bias is already a significant contributor to shear bias at LSST sensitivities, even for isolated galaxy images, shear calibration techniques have been recently developed to potentially remove pixel-noise bias in isolated galaxy images (see, for example, Bayesian Fourier domain (BFD)~\citep{2014MNRAS.438.1880B} and Metacalibration~\citep{2017arXiv170202600H, 2017ApJ...841...24S}). For blended objects, several methods that rely on image simulation campaigns or deep observations have been recently introduced~\cite{2015MNRAS.449..685H, 2017MNRAS.467.1627F, Samuroff2018, 2018MNRAS.481.3170M, 2019A&A...627A..59E, 2019A&A...624A..92K}, but they require careful selection in the deep observations~\cite{2018MNRAS.481.3170M}, and/or the simulations~\cite{2015MNRAS.449..685H, 2019A&A...627A..59E} used for calibration. Methods such as \textsc{METADETECT} are being developed~\citep{2020ApJ...902..138S} and successfully remove biases even at the level required by LSST analyses, while avoiding the need for image simulations or deep observations. However, for blends at different redshifts further calibration, such as that presented at ref.~\cite{2020arXiv201208567M}, is needed, and may require image simulations.

Achieving the potential sensitivity of the LSST data sample for cosmic shear measurements will require focused efforts such as large image simulation campaigns, or new deep observations~\cite{2019A&A...627A..59E}, and new ideas for object detection, measurement, fast image simulations, and possibly deblending. 

\acknowledgments{This work was supported by DOE Grants DE-SC0009920 (UC Irvine),  DE-SC009193 (University of Michigan), and DE-SC0019351 (Stanford), NSF Grant PHY-1404070 (Stanford), and LSST Corporation (I.M).  We thank the developers of the \textsc{GalSim} galaxy simulation package, and those contributing to the production of the LSST simulated catalog of stars and galaxies (\textsc{CatSim}) and the simulated observing strategy for a possible ten-year LSST survey (\textsc{OpSim}). 
This paper has undergone internal review in the LSST Dark Energy Science Collaboration. 
We thank our internal DESC reviewers Chihway Chang, Mike Jarvis, and Gary Bernstein for their helpful feedback.

D.P.K.\ wrote the initial version of the {\tt WeakLensingDeblending} simulation package, implemented the Fisher matrix calculations, and performed the analysis of $\Neffgrpf$ for galaxies.  
J.S.\ extended the $\Neffgrpf$ analysis to include the impact of stars and implemented the comparison of the predictions with the measured values from DES and HSC. 
I.M.\ wrote the software to calculate pixel-noise bias and performed the analysis of shape and cosmic shear bias due to pixel noise. 
D.P.K.\ and P.R.B.\ provided guidance throughout. 
All authors discussed interpretation of the results and contributed to writing the paper.

The authors acknowledge the usage of \texttt{numpy}, \texttt{scipy}, \texttt{astropy}, \texttt{lmfit}, \texttt{fitsio}, \texttt{corner}, \texttt{GalSim}, and \texttt{matplotlib}.

The LSST DESC acknowledges ongoing support from the Institut National de Physique Nucl\'eaire et de Physique des Particules in France; the Science \& Technology Facilities Council in the United Kingdom; and the Department of Energy, the National Science Foundation, and the LSST Corporation in the United States.  
The LSST DESC uses resources of the IN2P3 Computing Center (CC-IN2P3--Lyon/Villeurbanne - France) funded by the Centre National de la Recherche Scientifique; the National Energy Research Scientific Computing Center, a DOE Office of Science User Facility supported by the Office of Science of the U.S.\ Department of Energy under Contract No.\ DE-AC02-05CH11231; STFC DiRAC HPC Facilities, funded by UK BIS National E-infrastructure capital grants; and the UK particle physics grid, supported by the GridPP Collaboration. This 
work was performed in part under DOE Contract DE-AC02-76SF00515.
}

\def\apjl{ApJL} 
\def\aj{AJ} 
\def\apj{ApJ} 
\def\pasp{PASP} 
\def\spie{SPIE} %
\def\apjs{ApJS} 
\def\araa{ARAA} 
\def\aap{A\&A} 
\def\aaps{A\&A~Supl.} 
\def\nat{Nature} 
\def\nar{New Astron. Rev.} 
\def\mnras{MNRAS} 
\def\jcap{JCAP} 
\def\prd{{Phys.~Rev.~D}}        
\def\physrep{{Phys.~Reports}} 

\bibliographystyle{JHEP}
\bibliography{blending_journal}

\appendix

\section{Dependence of noise bias on galaxy size, ellipticity and shape parameter}
\label{app:dep_on_size_ell}

\begin{figure}[tbp]
\begin{center}
\includegraphics[width=0.95\textwidth]{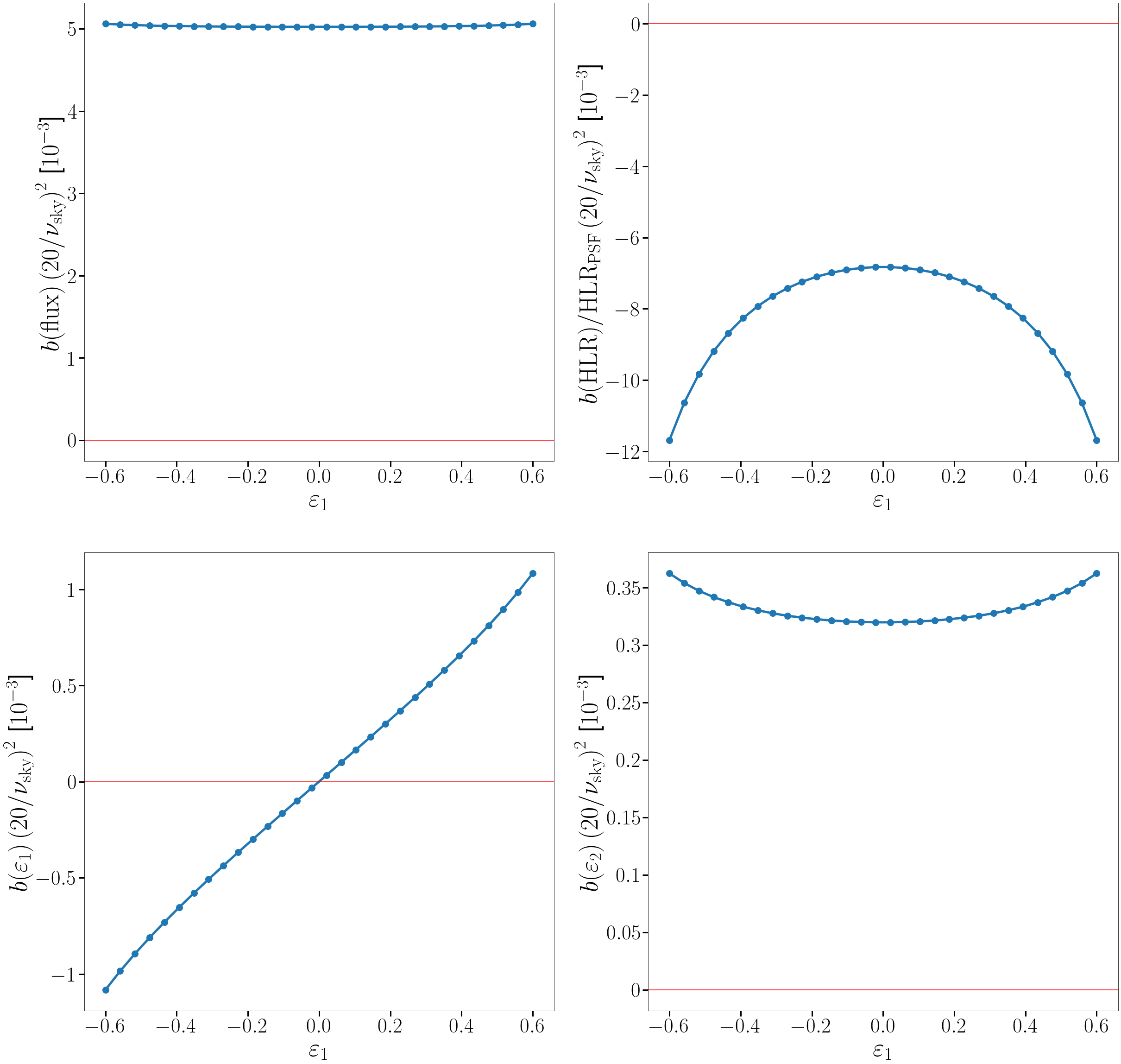}
\caption{Fisher predictions for dependence of noise bias $b(\varepsilon_1)$ on ellipticity component $\varepsilon_1$, for a single-galaxy image with parameter values ${\rm PSF\ HLR} = 0.35\,{\rm arcsec}$, ${\rm galaxy\ HLR} = 1.0\,{\rm arcsec}$, flux = 1 unit, and $\varepsilon_2 = +0.2$.}
\label{fig:single_gal_vary_e1}
\end{center}
\end{figure}

\begin{figure}[tbp]
\centering
\begin{subfigure}[b]{0.425\textwidth}
\includegraphics[width=1.0\textwidth]{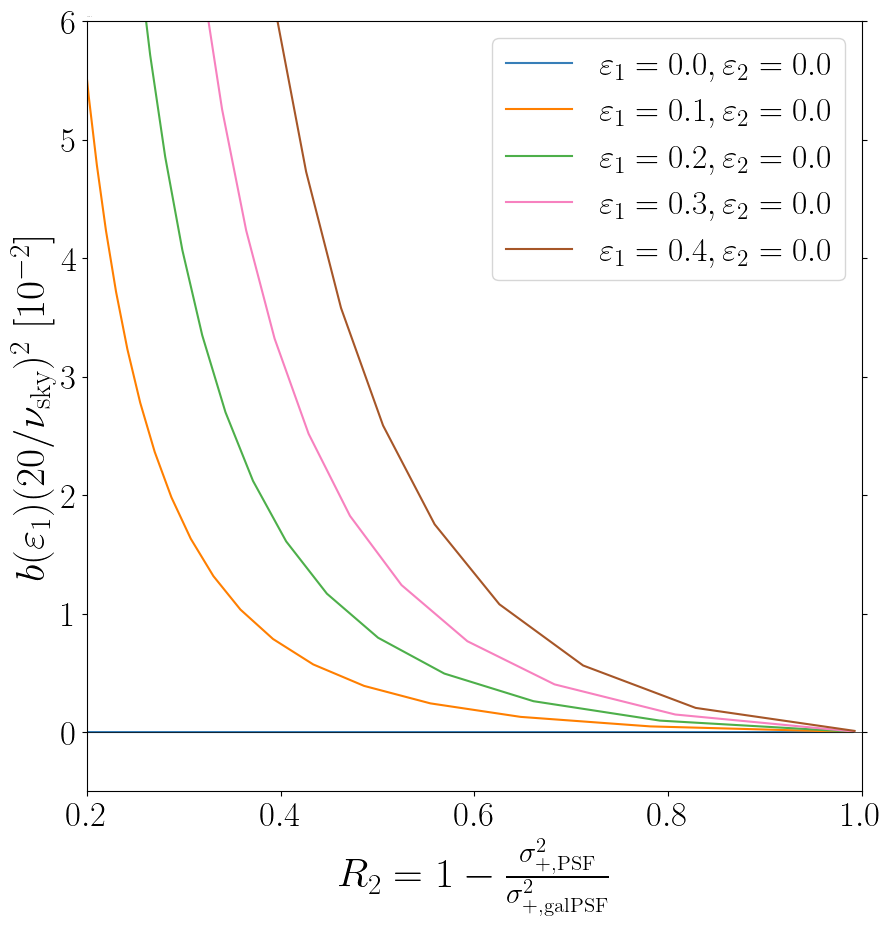}
\end{subfigure}
\quad
\begin{subfigure}[b]{0.45\textwidth}
\includegraphics[width=1.0\textwidth]{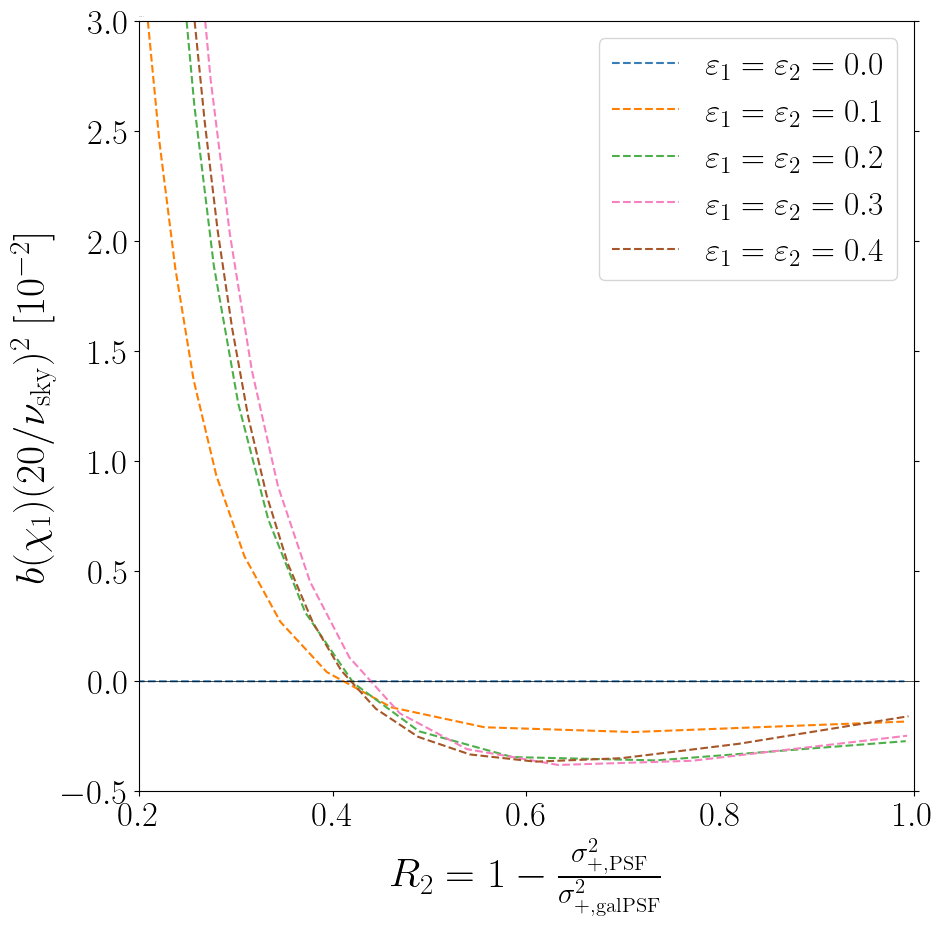}
\end{subfigure}
\caption{Fisher predictions for dependence of noise bias on the size parameter $R_{2}$ (defined in \eqn{definition-R2}) for ellipticity parameterizations $\varepsilon_1$ (left) and $\chi_1$ (right), for an isolated galaxy convolved with a PSF with HLR = 0.35 arcsec. This figure allows for direct comparison with Hirata et al. \cite{Hirata+M2004}.}
\label{fig:size_bias_eps2=0}
\end{figure}

We show in figures \ref{fig:single_gal_vary_e1} and \ref{fig:size_bias_eps2=0}
the Fisher predictions for noise bias as a function of several galaxy parameters, for an isolated galaxy. 
The nominal galaxy parameter values are ${\rm HLR_{PSF}} = 0.35\,{\rm arcsec}$ and ${\rm galaxy\ HLR} = 0.5\,{\rm arcsec}$ so that ${\rm HLR}/{\rm HLR_{PSF}} \sim 1.43$; flux = 1 unit, $\varepsilon_1 = \varepsilon_2 = +0.2$, and $\nu_{\rm sky} = 20$. For the vertical axis labels, we include a factor of  $(20/\nu_{\rm sky})^2$ to make explicit the Fisher prediction for the scaling of noise bias with SNR in the sky-dominated limit. 

In \fig{single_gal_vary_e1}, we see that the predicted bias on flux, HLR, and ellipticity components depends on the shape of the galaxy itself. 

In \fig{size_bias_eps2=0}, we compare the Fisher prediction for noise bias for two different commonly used shear estimators --  
$\epsv$, which was already introduced and defined in \eqn{ellipticity}, and  
$\chiv$, which differs only in the denominator:
\begin{equation}
\chiv \equiv \frac{Q_{11} - Q_{22} + 2 i Q_{12}}{Q_{11} + Q_{22}} \;.
\label{eqn:chi-ellipticity}
\end{equation}

Since $\epsv$ and $\chiv$ have different (nonlinear) dependences on the second moments $Q_{ij}$, and therefore pixel flux, we cannot expect the noise bias for each to be identical, for the same surface brightness profile and PSF.
We see in \fig{size_bias_eps2=0} that the sign and magnitude of the predicted bias depend sensitively on which shape estimator is used and the relative size of the galaxy and the PSF.
For both estimators, the bias tends to large positive values when the galaxy size is much smaller than the PSF size ($R_{2} \rightarrow 0$) because small changes in the estimated shape of the PSF-convolved image correspond to large differences in intrinsic ellipticity for poorly resolved galaxies.
%
\section{Source Extractor settings}
\label{app:sextractor_settings}
We use version 2.19.5 of Source Extractor. 
In \tab{sextractor_settings}, we list the ``default'' settings used to define classes of objects, as defined in sec.\,\ref{sec:sample_definitions}.
The convolution kernel (FILTER) is a Gaussian profile with FWHM equal to 2.0 pixels (0.4 arcsec) and represented by a $3\times 3$ array of pixel values. 

\begin{table}[tbp]
\centering
\begin{tabular}{|l l|} 
 \hline
Parameter & Value \\
 \hline
 DEBLEND\_NTHRESH & 32 \\ 
 DEBLEND\_MINCONT & 0.005 \\

 CLEAN\_PARAM & 1.0 \\
 BACK\_SIZE & 64 \\
 DETECT\_MINAREA & 5 \\
 DETECT\_THRESH & 1.5 \\
 \hline
\end{tabular}
\caption{Source Extractor parameter values used in this study.}
\label{tab:sextractor_settings}
\end{table}
\section{Additional estimates of SNR}
The \texttt{WeakLensingDeblending} package computes additional flux uncertainty $\sigma_{f}$ and SNR estimates beyond those presented in \tab{snr_defs}. We define these additional estimates in \tab{snr_defs_2}.

\begin{table}[tbp]
\begin{center}
\begin{tabular}{lcll}
Assumptions & Subscript & Assumed variance & Free parameters \\
\hline
Isolated source & iso & $b + s_{ip}$ & $f_i$ \\
Group & grp & $b + s_{ip} + s_{jp} + \ldots$ & $f_i$  \\
\hline
\end{tabular}
\caption{Definitions of the different Fisher-matrix uncertainty-estimation models used to define flux uncertainty $\sigma_f$ and SNR estimates $\nu = f/\sigma_f$. See caption of \tab{snr_defs} for details.}
\label{tab:snr_defs_2}
\end{center}
\end{table}

\section{\texorpdfstring{Plots of effective number density, $N_{\rm eff}$}{Neff}}
\label{app:Neff_plots}

\begin{figure}[h]
\begin{center}

\includegraphics[width=6.0in]{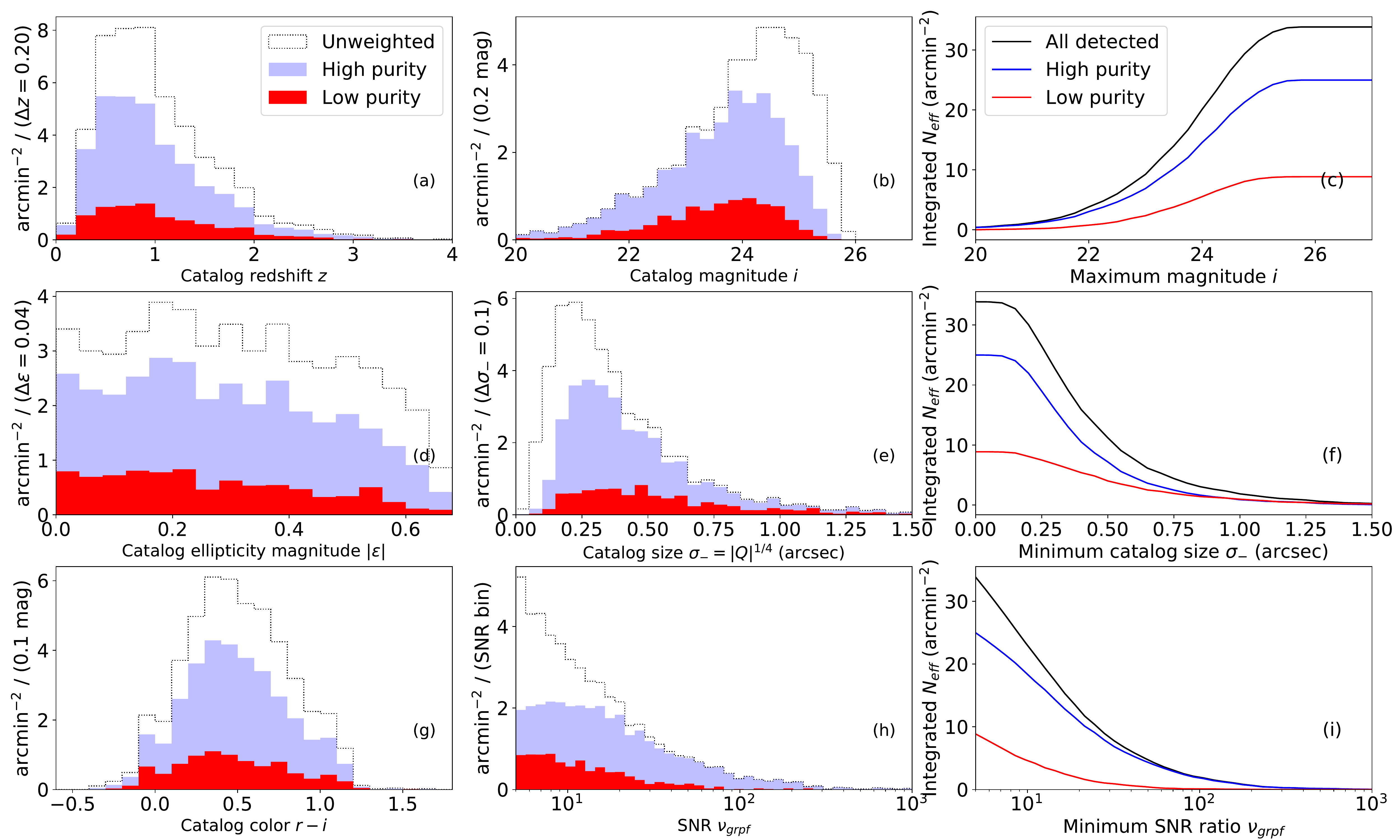}
\caption{Distributions of catalog parameters for detectable galaxies in a simulated HSC full-depth $i$-band exposure. See the caption to \fig{LSST_i} for details.}
\label{fig:HSC_i}
\end{center}
\end{figure}

\begin{figure}[h]
\begin{center}
\includegraphics[width=6.0in]{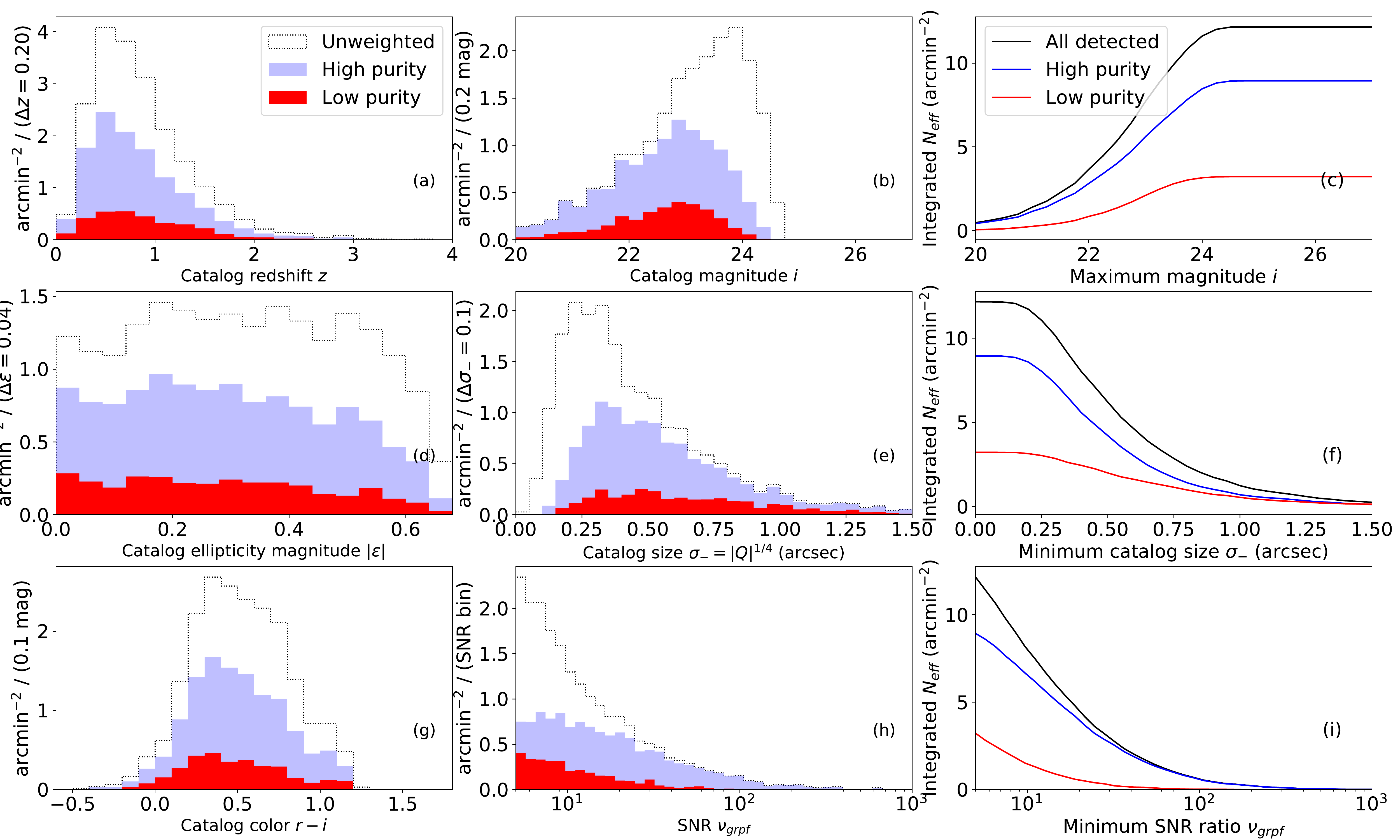}
\caption{Distributions of catalog parameters for detectable galaxies in a simulated DES full-depth $i$-band exposure. See the caption to \fig{LSST_i} for details.}
\label{fig:DES_i}
\end{center}
\end{figure}

In \fig{LSST_i} in  sec.\,\ref{sec:summary_predicted_Neff} (LSST $i$ band), and figures \ref{fig:HSC_i} (HSC $i$ band) and \ref{fig:DES_i} (DES $i$ band)
we show the distributions of six catalog parameters (redshift, $i$-band magnitude, ellipticity, size, color, $\nu_{\rm grpf}$) for detectable galaxies ($\nu_{\text{grpf}} > 6$). The dashed outline histograms show unweighted distributions while the filled red and blue regions show the cumulative weighted contributions of low-purity and high-purity galaxies, respectively, using the weights $w_i$ defined in \eqn{weights}. 
The right-hand plots show the integrated $\Neffgrpf$ as a function of (c) maximum magnitude in the simulated filter, (f) minimum galaxy size $\sigma_-$, and (i) minimum signal-to-noise estimate $\nu_{\text{grpf}}$.

\section{Details on the comparison with prior studies of \texorpdfstring{$N_{\rm eff}$}{Neff}}
\label{app:prior_studies}

In this section we describe in detail the comparison between our estimates of $\Neffgrpf$ and prior studies summarized in \sect{prior_studies}.

\subsection{Prior studies of \texorpdfstring{$N_{\rm eff}$}{Neff} for LSST}
 
In ref.~\cite{2013MNRAS.434.2121C}, Chang et al.\ use images generated with the  \texttt{PhoSim}~\cite{2015ApJS..218...14P} simulation package to predict $N_{\rm eff}$, which they define as 
\begin{equation}
N_{\rm eff}=\frac{1}{A}\sum_{i}\frac{\sigma^{2}_{I}}{\sigma^{2}_{I}+\sigma_{i}^{2}}\, ,
\label{eqn:chang_neff}
\end{equation}
where $\sigma_I = 0.26$ is their assumed intrinsic galaxy shape noise and the sum is over all galaxies in a field of area $A$ in the sample of interest. Their calculation uses Source Extractor~\cite{1996A&AS..117..393B} to determine which galaxies are detected and included in $\Neffgrpf$, which we emulate with the criterion $\nu_{\rm{grpf}} > 6$.

As in ref.~\cite{2013MNRAS.434.2121C}, we  model the shear measurement uncertainty $\sigma_{i}$ for galaxy $i$  as~\cite{Bernstein&J2002}
\begin{equation}
\sigma_{i}=\frac{a}{\nu_i}\left[1+\left(\frac{b}{R_i}\right)^{c}\,\right]\, , 
\label{eqn:sigma_fitting}
\end{equation}
where $R_i$ is the square of the ratio of the effective size (not convolved with the PSF) of the galaxy to the size of the PSF:
\begin{equation}
R_i=\frac{\sigma_{+,i}^{2}}{\sigma_{+,\, {\rm PSF}}^{2}}\,.
\end{equation}
The signal-to-noise ratio $\nu_i$ defined in both ref.~\cite{Bernstein&J2002} and ref.~\cite{2013MNRAS.434.2121C} is equivalent to our definition of $\nu_{\rm iso}$ from \tab{snr_defs_2}.
The authors of ref.~\cite{2013MNRAS.434.2121C} fit the RMS of the measured shear residuals (for a particular detection and measurement algorithm applied to simulated images) in bins of $R$ and $\nu$, obtaining $(a, b, c) = (1.58, 5.03, 0.39)$. 
It is important to note, as pointed out in ref.~\cite{2013MNRAS.434.2121C}, that the measured RMS is larger than the $\sigma$ parameter for the Gaussian core of the shear residuals distribution due to the small, but extended, non-Gaussian tails, which can be seen in figure 4 in ref.~\cite{2013MNRAS.434.2121C}). This fact will become relevant for understanding the differences between the results found in this work and in ref.~\cite{2013MNRAS.434.2121C}.

We compare our methodology with ref.~\cite{2013MNRAS.434.2121C} at two levels. First, we check whether our prediction for the shear measurement uncertainty $\sigma_{i}$ has the dependence on size and signal-to-noise ratio predicted by \eqn{sigma_fitting}. Second, we compare the predicted values for $\Neffgrpf$.

\begin{figure}
\centering
\includegraphics[width=0.75\textwidth]{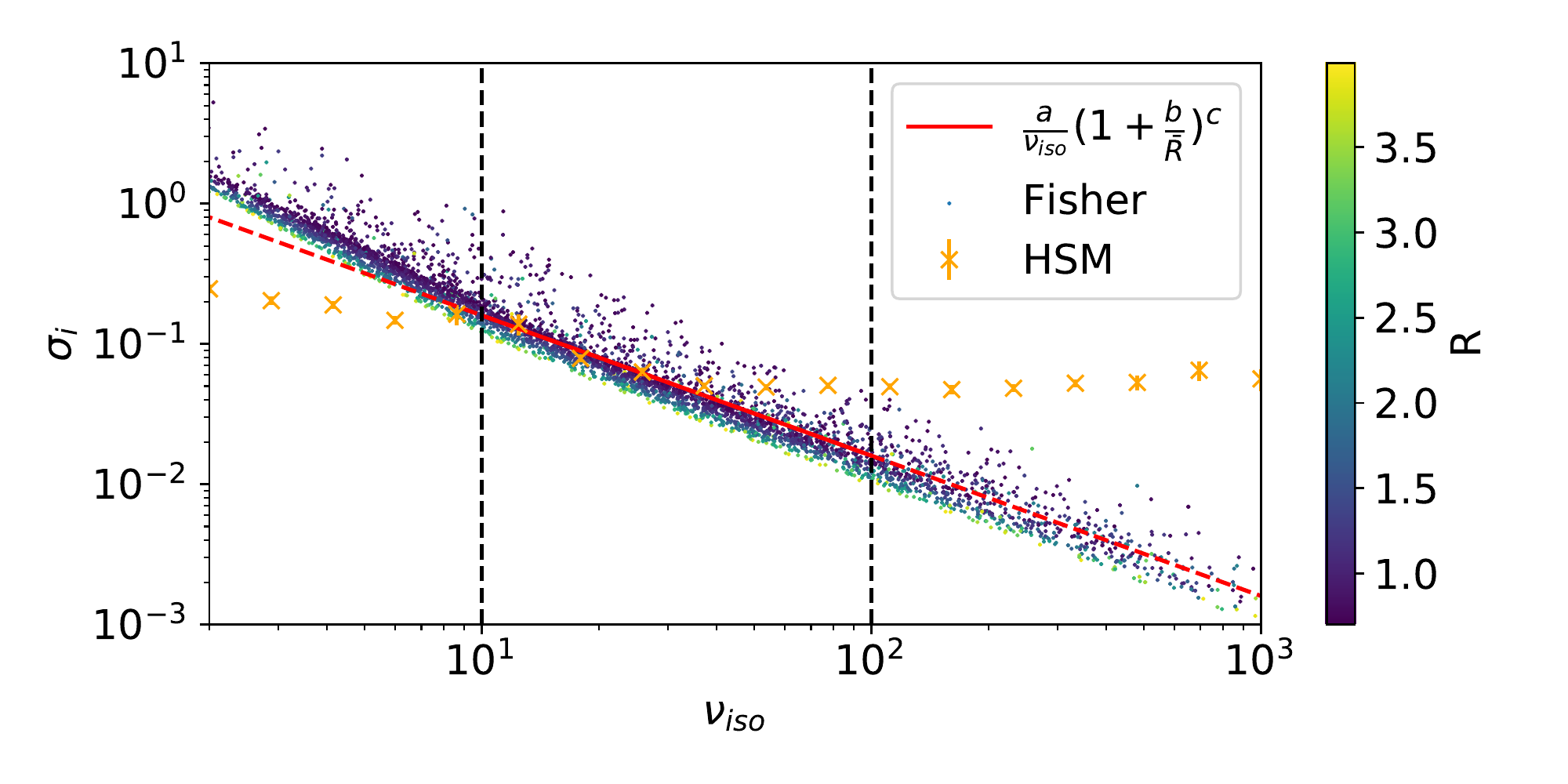}
\includegraphics[width=0.75\textwidth]{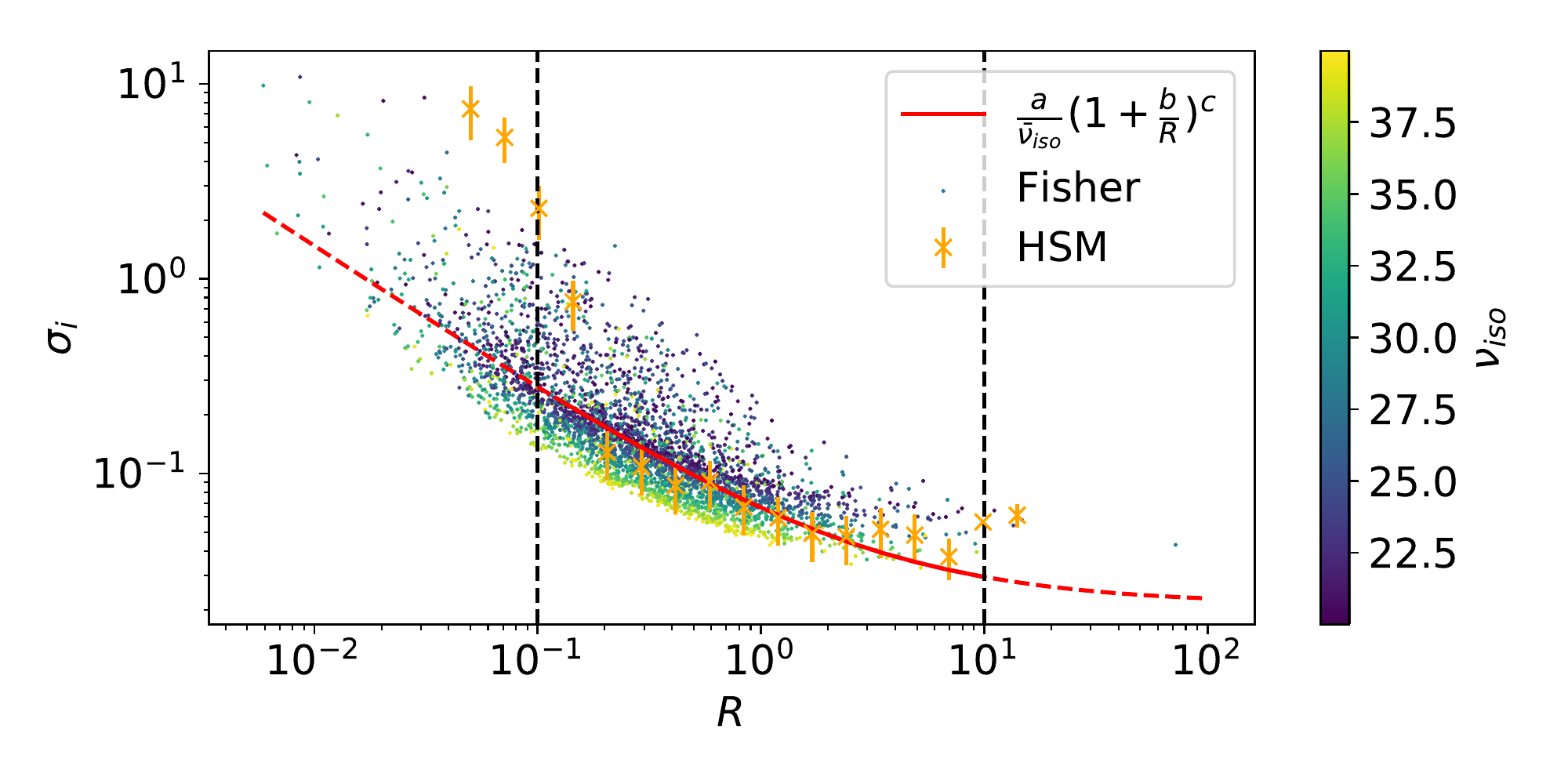}
\caption{Comparison of ellipticity measurement uncertainties for unblended galaxies in one simulated LSST chip in $r$ band. Colored dots correspond to the per-galaxy Fisher prediction for statistical uncertainty $\sigma_{i}$.  
The orange symbols ($\times$) correspond to the mean of the statistical uncertainties from the REGAUSS method of the HSM shape measurement package, with bootstrapped error bars, in 18 bins in the range $2 < \nu_{\rm iso} < 1000$ (top) and  $0.05 < R < 20$ (bottom). 
The top plot shows the evolution of these uncertainties with $\nu_{\rm iso}$ for galaxies with $ 0.7 < R < 4.0$, while 
the bottom plot shows the evolution with $R$ for galaxies with $ 20 < \nu_{\rm iso} < 40$. 
These ranges are chosen to illustrate two projections in the ($R$, $\nu_{\rm iso}$) space. The color of the dots indicates the value of $R$ (top) and $\nu_{\rm iso}$ (bottom). 
The red curve in each plot corresponds to the fit to \eqn{sigma_fitting} for $\sigma_{i}(\nu_{\rm iso}, R)$; the curves are solid in the range of $\nu_{\rm iso}$ and $R$ used for the fits (also indicated by the vertical dashed lines) and dashed elsewhere. These curves are evaluated for $R = 1.39$ (top) and $\nu_{\rm iso} = 28.09$ (bottom), which are the mean values of these parameters for the points shown in each plot.} 
\label{fig:error_comparison}
\end{figure}

For the measurement uncertainty $\sigma_{i}$, we use \eqn{effective_weight} where the component uncertainties $\sigma_{i,1}$ and $\sigma_{i,2}$ are obtained from the ``blending off'' Fisher matrix to be consistent with the definition of signal-to-noise ratio in ref.~\cite{2013MNRAS.434.2121C}. Then, we perform an unbinned least-squares fit to \eqn{sigma_fitting} of the predicted $\sigma_{i}$ considering only objects with  $0.1 < R < 10$, $10 < \nu_{\rm iso} < 100$, and $\sigma_i \leq 1$ ($\sim 10,000$ objects). We find parameter values of $a=0.60 \pm 0.13$, $b=2.7 \pm 1.2$, and $c=0.75 \pm 0.03$. 
Figure~\ref{fig:error_comparison} shows scatter plots of $\sigma_i$ (colored dots) as a function of $\nu_{\rm{iso}}$ for $0.7 < R < 4.0$ (top plot) and as a function of $R$ for $20 < \nu_{\rm{iso}} < 40$ (bottom plot); these ranges of $R$ and $\nu_{\rm{iso}}$ are chosen for illustration only. The best fit to \eqn{sigma_fitting} is shown as a red curve in each plot, with $\sigma_i(\nu_{\rm iso}, R)$ evaluated for $R = 1.39$ (top) and $\nu_{\rm iso} = 28.09$ (bottom), which are the mean values of these parameters for the points in each panel.  

We can see that \eqn{sigma_fitting} describes well the general trend of $\sigma_i$ with $\nu_{\rm{iso}}$ and $R$ for objects with $\nu_{\rm iso} > 10$. We find that the values for the parameters $(a, b, c)$ are different than those found in ref.~\cite{2013MNRAS.434.2121C}. This difference is due to the fact that, as mentioned earlier, the fits in ref.~\cite{2013MNRAS.434.2121C} were made to the RMS of the distribution of the shape residuals, whereas in this work we use the Fisher prediction for $\sigma_i$, which roughly corresponds to the Gaussian core of that distribution. More precisely, our Fisher formalism provides the optimal $\sigma_i$ for an unbiased estimator.

As further validation of our ellipticity measurement uncertainties, we calculate the average difference between the measured and true ellipticities in bins of $\nu_{\text{iso}}$ and $R$ (see \fig{error_comparison}),
where the measurements are performed on unblended images, including noise, with the REGAUSS method of the HSM shape measurement package~\cite{2003MNRAS.343..459H, Hirata&S2004}
in \textsc{GalSim}~\cite{2015A&C....10..121R}. 
The ellipticities are computed with ``adaptive'' moments that use an elliptical weight function, the shape of which is matched to that of the object. The measured moments of the galaxy are corrected for the PSF to produce the ellipticity of the intrinsic galaxy image \citep{Bernstein&J2002}. We see in \fig{error_comparison} that the REGAUSS uncertainties follow the Fisher predictions for 
$10 \lesssim \nu_{\rm{iso}} \lesssim 50$, as expected since this is the SNR regime for which HSM is optimized (Mandelbaum, private comm.). However, for $\nu_{\rm iso} < 10$ only about 25\% of REGAUSS measurements converge, resulting in an average uncertainty of the converged fits that is biased low.  In the high SNR regime, $\nu_{\rm iso}>100$, REGAUSS uncertainties are much larger than the Fisher predictions because the contributions of the tails of large galaxies are not fully captured by REGAUSS (Mandelbaum, private comm.).
  
Our second check is to compare our estimations of $\Neffgrpf$ with ref.~\cite{2013MNRAS.434.2121C} in which three scenarios to select sources are considered: $\sigma_i < k \sigma_I$ with $k=2$ (``optimistic"), $k=1$ (``fiducial"), or $k=0.5$ (``pessimistic"). The value of $k$ roughly parametrizes three different SNR regimes where shear measurement algorithms are found to have quantitatively different performance~\cite{2013MNRAS.434.2121C}. To emulate these scenarios in our framework, we always require $\nu_{\rm grpf}>6$ as a proxy for detection and use the Fisher-predicted values for $\sigma_i$ in \eqn{chang_neff}. 
We calculate an instrinsic shape noise of $\sigma_I = 0.24$ for the detected galaxies, somewhat lower than the value $\sigma_I = 0.26$ assumed in ref.~\cite{2013MNRAS.434.2121C}. Table~\ref{tab:comparison_chang} summarizes our results for each of these values. 
We would expect our calculated values of $\Neffgrpf$ to be most directly comparable to those in  ref.~\cite{2013MNRAS.434.2121C} for $\sigma_I = 0.26$, but slightly lower since we include recent updates to the LSST throughput model\footnote{\url{https://github.com/lsst-pst/syseng_throughputs/archive/1.4.zip}} that reduce the overall throughput by $\sim15\%$,  lowering the total number of ``detectable'' objects. However, as discussed earlier in this section, our Fisher predicted values for $\sigma_{i}$ are typically smaller than those found in ref.~\cite{2013MNRAS.434.2121C}, compensating for this loss in ``detectable" objects and resulting in larger values for $\Neffgrpf$. This is especially significant for the scenarios with $k=0.5$ and $k=1.0$. As mentioned earlier, in ref.~\cite{2013MNRAS.434.2121C} $\sigma_{i}$ was estimated from the RMS of the residual ellipticity distribution, which leads to larger values than the Fisher-predicted $\sigma_{i}$ since the latter are more closely related to the size of the Gaussian core of that distribution. The lower values of $\sigma_{i}$ for the Fisher analysis lead to higher weights for each galaxy and a higher fraction of galaxies meeting the selection criteria, resulting in larger values of $\Neffgrpf$. For the optimistic scenario, more low-SNR galaxies are selected; however, these low SNR galaxies will contribute fractionally less to the overall value of $\Neffgrpf$ and the discrepancy with ref.~\cite{2013MNRAS.434.2121C} decreases.

\begin{table}
\centering
\begin{tabular}{|cccc|}
\hline
$\Neffgrpf$ estimator & Pessimistic $(k=0.5)$ & Fiducial $(k=1.0)$ & Optimistic $(k=2.0)$\\
\hline
This work, $\sigma_I = 0.24$ & 28.9 & 37.2 & 41.0\\
This work, $\sigma_I = 0.26$ & 30.0 & 38.4 & 42.3\\
Chang et al., $\sigma_I = 0.26$ & 20 & 30 & 39\\
\hline
\end{tabular}
\caption{Summary of estimated effective number density $\Neffgrpf$ in units of galaxies/arcmin$^{2}$ for the value of $\sigma_{I} = 0.24$ from our ``detectable" sample, and $\sigma_{I} = 0.26$ as in ref.~\cite{2013MNRAS.434.2121C}. Note that $\sigma_{I}$  depends on the sample; therefore, after we select the detected galaxies with $\sigma_{i} < \sigma_{I}$, the new sample has a different value for $\sigma_{I}$.}
\label{tab:comparison_chang}
\end{table}

\subsection{Prior studies of \texorpdfstring{$N_{\rm eff}$}{Neff} with the Dark Energy Survey}

We compare the value of $\Neffgrpf$ that we calculate for the Dark Energy Survey with \eqn{Neff_weights} to the results published by the DES collaboration for Science Verification data~\cite{2016MNRAS.460.2245J}. 
We apply the following selection criteria to approximate the selection used in the DES  \texttt{NGMIX} analysis described in ref.~\cite{2016MNRAS.460.2245J}:
$R>0.15$,
$\nu_{\rm sky}>15$,
$18 < i < 23$ where $i$ is the $i$-band magnitude,\footnote{The cut $i < 23$ was made to match the limiting magnitude of the DES-SV Gold catalog:  \url{https://des.ncsa.illinois.edu/releases/sva1/doc/gold}. We used an exposure time similar to that of DES-SV: 1000 s in $i$ band.} and
$i + 3.5\log{\left(\frac{f_{i}}{2\sigma_{+}^{2}}\right)} < 18$, where $f_{i}$ is the flux in the $i$ band. 

If we use the criterion $\nu_{\rm grpf}>5$ as a proxy for detection, we obtain $\Neffgrpf=5.8$ galaxies/arcmin$^{2}$, in good agreement with the value of $N_{\rm eff}=5.7$ galaxies/arcmin$^{2}$ found in ref.~\cite{2016MNRAS.460.2245J}.

\subsection{Prior studies of \texorpdfstring{$N_{\rm eff}$}{Neff} with Subaru HSC-SSP}

We compare the value of $\Neffgrpf$ that we calculate with \eqn{Neff_weights} for a sample similar to the Subaru HSC-SSP survey year 1 shear catalog to the results published by the HSC collaboration~\cite{2018PASJ...70S..25M}. 
We apply the following selection criteria to approximate that used in the HSC analysis described in ref.~\cite{2018PASJ...70S..25M}:
\begin{itemize}
\item $\nu_{\rm sky} > 10$, 
$i < 24.5$, and  $\sqrt{\frac{1}{2}\left(\sigma^{2}_{i,1}+\sigma_{i,2}^{2}\right)} < 0.4$;
\item $|\pmb{\varepsilon^{\prime}}_{\rm HSM}|^{2}<2$, with $\pmb{\varepsilon^{\prime}}_{\rm HSM}$ defined as the measured PSF-corrected ellipticity vector measured by the REGAUSS method in HSM; 
 
\item $R_{2} \geq 0.3$, where $R_{2}$ is a resolution factor defined as
\begin{equation}
R_{2}=1-\frac{\sigma^{2}_{+,{\rm PSF}}}{\sigma^{2}_{+,{\rm galPSF}}},
\label{eqn:definition-R2}
\end{equation}
where $\sigma^{2}_{+,{\rm galPSF}}$ is the size of the PSF-convolved galaxy image as defined in \eqn{size}.
\item Blendedness $ < 10^{-0.375}$, which is equivalent to purity $\rho > 0.578$.
\end{itemize} 

With the above critera applied, we  use  \eqn{Neff_weights} to find  $\Neffgrpf=22.3$ for galaxies with $\nu_{\rm grpf}>5$ detection threshold. 
This value is  very close to the result  found  by HSC in ref.~\cite{2018PASJ...70S..25M}:  $N_{\rm eff}=21.8$ galaxies/arcmin$^{2}$.

As mentioned earlier, we expect our estimated $\Neffgrpf$ -- with the same purity threshold as the data -- to be higher than the $N_{\rm eff}$ values in HSC data due to the optimistic nature of our lensing weights and possible observational effects missing in our simulation. 
However, the results are remarkably close after following the selection procedure, giving confidence in the robustness of our methodology.
\end{document}